\documentclass[letterpaper,11pt,nofootinbib]{revtex4-2}
\pdfoutput=1
\usepackage{xcolor}
\usepackage{amsmath,amsthm,amsfonts,amssymb,braket}
\usepackage[pdftex,colorlinks=true,linkcolor=blue,citecolor=darkred,urlcolor=darkblue]{hyperref}
\usepackage[all]{xy}
\usepackage{enumitem}	
\usepackage{fullpage}
\usepackage{mathtools}
\usepackage{cleveref}
\usepackage{algorithm}

\usepackage{tikz-cd}

\newtheorem{lemma}{Lemma}[section]
\newtheorem{theorem}[lemma]{Theorem}

\newtheorem{proposition-definition}[lemma]{Proposition-Definition}
\theoremstyle{definition}

\newcommand{\nslice}{{n_{\textrm slice}}}

\newcommand{\RR}{{\mathbb{R}}}
\newcommand{\ZZ}{{\mathbb{Z}}}

\newcommand{\vol}{\textrm{vol}}

\newcommand{\sys}{{\rm sys}_{1,1}}
\newcommand{\systwo}{{\rm sys}_{1,2}}

\DeclarePairedDelimiter{\norm}{\lVert}{\rVert}
\DeclarePairedDelimiter{\abs}{|}{|}

\definecolor{darkblue}{rgb}{0.,0.,0.4}
\definecolor{darkred}{rgb}{0.5,0.,0.}
\definecolor{darkpurple}{rgb}{0.5,0.,0.5}
\definecolor{ltgreen}{rgb}{0.1,.59,.43}
\definecolor{orange}{rgb}{1.0, 0.5, 0.0}

\makeatletter
\def\l@subsubsection#1#2{}
\makeatother
\newcommand{\nocontentsline}[3]{}
\newcommand{\tocless}[2]{\bgroup\let\addcontentsline=\nocontentsline#1{#2}\egroup}

\newcommand{\jh}[1]{}
\newcommand{\mh}[1]{}
\newcommand{\da}[1]{}
\newcommand{\dave}[1]{}

\begin{document}
\title{Geometrically Enhanced Topological Quantum Codes}
\author{David Aasen$^*$}
\author{Jeongwan Haah$^{*,\dagger}$}
\author{Matthew Hastings$^*$}
\author{Zhenghan Wang$^*$}

\begin{abstract}
We consider geometric methods of ``rotating" the toric code in higher dimensions to reduce the qubit count. 
These geometric methods can be used to prepare higher dimensional toric code states using single shot techniques, and in turn these may be used to prepare entangled logical states such as Bell pairs or GHZ states.  This bears some relation to measurement-based quantum computing in a twisted spacetime.  We also propose a generalization to more general stabilizer codes, and we present computer analysis of optimal rotations in low dimensions.
We present methods to do logical Clifford operations on these codes using crystalline symmetries and surgery, and we present a method for state injection at low noise into stabilizer quantum codes generalizing previous ideas for the two-dimensional toric code.
\end{abstract}

\maketitle

\tableofcontents

\section{Introduction}
While \emph{topological} codes such as the toric code are standard in quantum computing, the \emph{geometry} of these codes can be used for practical advantages.  As a well-known example,
the two-dimensional toric code on an $d$-by-$d$ torus
gives a $[[2d^2,2,d]]$ code.  However, by rotating
the torus by $45$ degrees\cite{wen2003quantum,bombin2007optimal}, one can construct a $[[d^2,2,d]]$ code for \emph{even} $d$.  This rotation can be understood as cellulating the torus by taking the vertices at integer points modulo a lattice $\Lambda$ who basis vectors are rows of the matrix
$$\begin{pmatrix}
    d/2 & d/2 \\ -d/2 & d/2
\end{pmatrix}.$$
The number of vertices (and hence one-half the number of physical qubits) is equal to the determinant of this matrix, while the distance $d$ of the code is the minimum $\ell_1$ norm of a nonzero vector in $\Lambda$.
In this paper, we use this generalizations of this kind of lattice form to investigate rotated lattices in higher dimensions where a variety of effects can occur and large savings can be found.  To give a simple example of the advantage of this lattice method for understanding rotations in \emph{two} dimensions, note that it lets us immediately write down an optimal $[[d^2+1,2,d]]$ rotated toric code with \emph{odd} distance\footnote{See also \cite{kovalev2013quantum} for such a code.} using the lattice generated by the rows of
$$\begin{pmatrix}
    (d+1)/2 & (d-1)/2 \\ -(d-1)/2 & (d+1)/2
\end{pmatrix}.$$  

Using higher-dimensional generalizations of such lattices, we also propose an alternative architecture for a quantum computer.
Almost all proposed architectures for a quantum computer rely on first implementing Clifford operations in a fault tolerant fashion, and then using some methods (such as magic state distillation) to implement non-Clifford operations on top.  The standard methods for implementing these Clifford operations are to use some stabilizer code (e.g., the surface code, but other codes are also proposed) to encode qubits and protect them against error, and then implement logical operations between these stabilizer codes by a various methods (e.g., transversal CNOT gates, lattice surgery, or other methods).

Here we propose a more direct architecture.  One creates stabilizer states by preparing all physical qubits in some product state and then measuring stabilizers of some code which admits single shot error correction~\cite{campbell2019theory,bombin2015single}.  Then, we propose to ``slice" these stabilizer states by measuring out some subset of the physical qubits.  This is done in such a way that the remaining qubits form some number, $\nslice\geq 1$, of distinct stabilizer codes, with entanglement between the logical qubits of these codes.  This gives a direct route to produce several stabilizer codes, whose logical qubits are themselves in some entangled stabilizer state.  In a case analyzed below, we slice a three-dimensional toric code into several two-dimensional toric codes, allowing one to produce a Bell pair or higher GHZ state. In addition, we give a prescription to produce arbitrary encoded stabilizer states in~\cref{slicetwist}.  In~\cref{pathtocliffords}, we then explain how to build a Clifford computer from this.

This method bears some relation to measurement-based quantum computing~\cite{briegel2009measurement}.  However, we propose novel methods for creating entangled logical states.  Further, we reduce qubit count significantly by exploiting geometry using the lattice idea above; this idea in three or four dimensions is related to the idea of \cite{hastings2017quantum} in asymptotically large dimensions.  For example, we rotate the boundary conditions of the three-dimensional toric code to reduce the qubit count for the given code distance, as explained in~\cref{geometrymethod}.  This is a special case of the more general method in~\cref{slicetwist}.

In section \cref{geometrymethod}, we present the basic idea of rotated lattices in higher dimensions and give some examples.
In \cref{pathtocliffords}, we show how to create logical Bell pairs and use this as an architecture for a Clifford computer;
this section introduces the idea of slicing, where a lattice in some number of dimensions is ``sliced" into lower dimensional lattices, and correspondingly a higher dimensional code in a fixed logical state is ``sliced" into multiple lower dimensional codes in an entangled state.
In \cref{crystallinegates}, we explore symmetries of these codes that allow us to implement logical Clifford gates on the qubits of a single block.
In \cref{surgerysection} and \cref{injectionsection}, we further present methods of implementing Clifford operations by surgery and state injection.  The state injection methods are general and may be applied to arbitrary quantum stabilizer codes; they generalize ideas in \cite{li2015magic}.
The sections up to \cref{injectionsection} are largely focused on rotated toric codes in modest dimensions (roughly, three through five).
Then, in \cref{slicetwist} we generalize slicing to more general CSS codes; in \cref{higherdim}, we consider some of the asymptotics of this rotation, introducing an ``$\ell_1$ Hermite constant"; and in \cref{24cell}, we consider alternate cellulations other than the hypercubic lattice.
These sections \cref{slicetwist,higherdim,24cell} present more mathematical generalizations beyond the specific cases considered in earlier sections of this paper.
Finally, \cref{numsim} describes numerical techniques used to study these codes.

\section{A method to produce geometric codes with enhanced code performance}
\label{geometrymethod}

We introduce a method for reducing the qubit overhead for translation invariant codes defined on a $d$-tori. 
In the following subsections we analyze the simplest three-dimensional and four-dimensional examples coming from taking a three or four-dimensional toric code. 
The qubit reduction is a generalization of the usual ``rotation" that has been used in the two-dimensional toric code.
Such rotations can be performed at minimal cost to the code distance, but in higher dimensions can have striking reductions in the code parameters. 
For a large system, the qubit count grows with the volume of the torus, and this geometric enhancement begins to have diminishing returns. 


\subsection{Rotated toric code and Hermite normal form}

We consider a standard $n$-dimensional toric code:
take the simple (hyper)cubic lattice,
and put qubits on cells of a fixed dimension, $X$-checks on one lower dimensional cells, 
and $Z$-checks on one higher dimensional cells.
Suppose we have an integral lattice $\Lambda$,
a subgroup of the abelian group~$\RR^n$ under addition where each element of~$\Lambda$ has integral coordinates.
We take the toric code on $\RR^n/\Lambda$, with degrees of freedom on edges (i.e., $1$-cells).
If $M$ is a matrix containing basis vectors of~$\Lambda$ in its rows,
then the volume $\vol(\RR^n / \Lambda)$ is equal to $\abs{\det M}$,
and the total number of qubits is the product of this volume of the torus and the number of qubits per unit hypercube.
For this $(1,n-1)$-toric code, the code distance is given by the $\ell_1$ $1$-systole.
That is, it is the minimum $\ell_1$ norm of a nonzero vector in the lattice.  We denote this by $\sys$, with the subscripts indicating that this is the $1$-systole computed using the $\ell_1$ norm.

It will often be convenient to bring $M$ to so-called Hermite normal form, by left multiplying by a unimodular matrix which leaves the lattice invariant.  In this form, $M$ is an upper triangular matrix, that is:
\begin{align}
    i<j \; \Longrightarrow \; 0\leq M_{ij} < M_{jj}.
\end{align}
These conditions define the Hermite normal form.
This form has two advantages for us.  For one, it simplifies a computer search for lattices with minimum determinant for the given $\ell_1$ systole, as the determinant can be immediately calculated in this form (being simply the product of diagonal entries).  Second, in our construction of ``easily sliceable" codes below (see \cref{slicing}), we can immediately read off whether a given three-dimensional toric code can be ``sliced" into two or more two-dimensional toric codes by examining the upper left diagonal entry of the Hermite normal form.  Suppose the upper left entry is equal to some integer $\nslice>1$.  Then, we may divide the qubits of the toric code into $\nslice$ disjoint sets. Label the directions of space $1,2,3$ corresponding to the three columns of the matrix in Hermite normal form.  
One set of qubits are those corresponding to edges pointing in direction $1$.  Once these qubits are measured in the $X$-basis (see \cref{slicing}), the vertex stabilizers of the three-dimensional toric code become vertex stabilizers of $\nslice$ distinct two-dimensional toric codes.  Remark: depending on the measurement outcome, there may of course be some sign in front of the vertex stabilizers of the two-dimensional toric code, but we continue to refer to it as a two-dimensional toric code state as it is the same up to a Pauli frame change.  Each of these two-dimensional toric codes corresponds to one of the $\nslice$ possible choices of the first coordinate in the three-dimensional geometry.

\subsection{Examples in three dimensions}
We give some minimal examples in three dimensions, i.e.,~the minimal possible determinant for given $\ell_1$ systole.
Every example generates a family of examples: for any integer $\ell\geq 1$ we may multiply all entries of the matrix $M$ by $\ell$, increasing the $\ell_1$ systole by a factor of $\ell$ and increasing the determinant by a factor of $\ell^3.$

The results are summarized in \cref{disttablenoslice} if we do not require that it be easily sliceable, and in \cref{disttableslice} when we require $\nslice\geq 2$.  In fact, in every case, the minimal determinant was found with the minimal number of slices; i.e, for given $\ell_1$ systole, the minimum determinant with no requirement on $\nslice$ was found with $\nslice=1$ and the minimum determinant requiring $\nslice\geq 2$ was found with $\nslice=2$.

\begin{table}[h!]
\caption{Table of codes without requiring that it be easily sliceable.
In the first table, the first column is the $\ell_1$ systole, the second column is the determinant of the lattice, and the third column is the ratio of the determinant to the cube of the $\ell_1$ systole.
The second table contains example Hermite normal forms (HNF) that realizes the code distances.
}
\begin{tabular}{c|l|l}
\hline
$\sys$ & $\det$ & $\det/\sys^3$\\
\hline
2 & 2 & 0.25\\
3 & 7 & 0.259\ldots \\
4 & 12 & 0.1875\\
5 & 27 & 0.216\\
6 & 38 & 0.1759\ldots\\
7 & 70 & 0.204\ldots\\
\hline
\end{tabular}
\quad\quad
\begin{tabular}{c|ccc}
$\ell_1$ systole     &  $3$ & $5$ & $7$ \\
\hline
HNF  & $\begin{pmatrix}
        1 & 0 & 4 \\
        & 1 & 5 \\
        & & 7
    \end{pmatrix} $
    &
    $\begin{pmatrix}
        1 & 0 & 17 \\
        & 1 & 23 \\
        & & 27
    \end{pmatrix}$
    &
    $\begin{pmatrix}
        1 & 0 & 45 \\
        & 1 & 54 \\
        & & 70
    \end{pmatrix}$
\end{tabular}
\label{disttablenoslice}
\end{table}





\begin{table}[h!]
\caption{Table of codes with $\nslice\geq 2$.  
In the first table, the first column is the $\ell_1$ systole, the second column is the determinant of the lattice, and the third column is the ratio of the determinant to the cube of the $\ell_1$ systole.
The second table contains example Hermite normal forms (HNF) for odd code distances.}
\begin{tabular}
{c|l|l}
\hline
$\sys$ & $\det$ & $\det/\sys^3$\\
\hline
2 & 4 & 0.5\\
3 & 10 & 0.37\ldots\\
4 & 16 & 0.25\\
5 & 30 & 0.24\\
6 & 44 & 0.2037\ldots\\
7 & 72 & 0.2099\ldots\\
\hline
\end{tabular}
\quad\quad
\begin{tabular}{c|ccc}
$\ell_1$ systole     &  $3$ & $5$ & $7$ \\
\hline
HNF  & $\begin{pmatrix}
        2 & 0 & 4 \\
        & 1 & 3 \\
        & & 5
    \end{pmatrix} $
    &
    $ \begin{pmatrix}
        2 & 0 & 9 \\
        & 1 & 11 \\
        & & 15
    \end{pmatrix}$
    &
    $\begin{pmatrix}
        2 & 1 & 5 \\
        & 3 & 8 \\
        & & 12
    \end{pmatrix} $
\end{tabular}
\label{disttableslice}
\end{table}




As we will explain in~\cref{errorcorrect} below,
certain error modes have effective distance 
that is twice the $\ell_1$ systole.
Exploiting this, we may further reduce the qubit count
while maintaining the effective code distance in regards to Bell state generation.
We give such examples in~\cref{tb:sliceable-shallow}.

\subsection{Higher dimensions}
\label{fourandhigher}
In four dimensions, in this subsection we again consider toric codes on hypercubic lattices.  Now we may consider two different codes, either a $(1,3)$ or a $(2,2)$ toric code.  In the $(2,2)$ toric code on a hypercubic lattice, stabilizers have weight $6$.  In \cref{24cell}, we consider other cellulations of a four-torus to reduce stabilizer weight.
In this section we primariy focus on the lattices and codes resulting from a numerical search (see \cref{numsim} for details) finding the optimal $\det$ for given $\sys$.

First, without imposing that it is easily sliceable, see \cref{disttable4d} and let us go through the examples one-by-one.  The column $\det(L)$ gives the minimum possible determinant for the given $\sys(L)$ for the given lattice $L$.  In some cases, as discussed below, the lattice with that $\det(L)$ and $\sys(L)$ is unique up to various symmetries.
\begin{table}[h!]
\caption{Table of parameters for various codes.  The ratio $\frac{n}{(2,2) \textrm{distance}^2}$ is the ratio of number of qubits to code distance squared.  For the $\det=16$ code, we have used $9$ for the distance, in the case that this is treated as a subsystem code.}
\begin{tabular}{l|l|l|l|l|l}
\hline
$\sys(L)$ & $\det(L)$ & $\frac{det(L)}{\sys(L)^4}$ & $(2,2)$ distance & $\frac{n}{(2,2) \textrm{distance}^2}$&comments \\
\hline
2 & 2 &$ \frac{1}{8}=0.125$ & 2 &3&$D_4$ lattice\\
3 & 9 & $\frac{1}{9}=0.111\ldots$ & 6& 1.5 & \\
4 & 16 & $\frac{1}{16}=0.0625$ & 8 &1.185\ldots& distance $9$ as subsystem code \\
5 & 45 & $\frac{45}{5^4}=0.072$ & 15 &1.2&\\
6 & 68 & $\frac{68}{6^4}=0.052\ldots$ &&& \\
7 & 152 & $\frac{152}{7^4}=0.063\ldots$ &&&\\
\hline
\end{tabular}
\label{disttable4d}
\end{table}

The column $(2,2)$ distance gives the distance of these examples as a $(2,2)$ toric code.  In some, but not all, of the cases where the lattice is not unique, we have checked the $(2,2)$ distance of some other lattices with the given $\sys(L)$ and $\det(L)$ and found that it is not as large as in the example here.  However, we have not performed an exhaustive search to optimize the $(2,2)$ distance for the given $\det(L)$.

For $\sys=2$, the optimal example has Hermite normal form
$$
\begin{pmatrix}
1 & 0 & 0 & 1\\
0 & 1 & 0 & 1\\
0 & 0 & 1 & 1\\
0 & 0 & 0 & 2
\end{pmatrix}.$$
This gives the so-called $D_4$ lattice.  It is the lattice of points with integer coordinates such that the sum of all coordinates is even.

For $\sys=3$, an optimal example has Hermite normal form
$$
\begin{pmatrix}
1 & 0 & 0 & 5\\
0 & 1 & 0 & 6\\
0 & 0 & 1 & 7\\
0 & 0 & 0 & 9
\end{pmatrix}.$$
Among the lattices with $\sys(L)=3$ and $\det(L)=9$, with the given entries on the diagonal of the Hermite normal form, this example is unique up to a permutation of rows and columns and up to changing the sign of the three off-diagonal elements in the last column mod $9$.
That is, we can make any of the exchanges $5\leftrightarrow -5=4\mod 9$, $6\leftrightarrow -6=3\mod 9$, or $7\leftrightarrow -7=4\mod 2$ in the last column and get a lattice with the same $\sys(L)$ and $\det(L)$.  These exchanges correspond to reflecting the lattice about a hyperplane.  We can also permute the three off-diagonal elements in the last column, which corresponds to a permutation of rows and columns (a column permutation is a combination of a rotation and reflection of the lattice, while a row permutation is a change of basis).

There are also examples with the same $\sys(L)=3$ and $\det(L)$ where the Hermite normal form has diagonal entries $1,1,3,3$.  However these lattices have smaller $(2,2)$ distance.

For $\sys=4$, we find a similar result.  The optimal $\det$ is $16$, and among lattices with Hermite normal form having diagonal elements $1,1,1,16$, an optimal
example is
$$
\begin{pmatrix}
1 & 0 & 0 & 7\\
0 & 1 & 0 & 5\\\
0 & 0 & 1 & 3\\
0 & 0 & 0 & 16
\end{pmatrix}.
$$
Again, with the given diagonal entries, this lattice is unique up to permuting the three off-diagonal entries in the last column and changing their sign.
There are also lattices with the same $\sys(L)$ and $\det(L)$ but different diagonal entries, but so far as we have tested these have a worse $(2,2)$ distance.

The $(2,2)$ distance displays an interesting phenomenon.  There is a nontrivial logical operator of weight $8$.  Indeed, considering just $Z$-type logical operators, we find $8$ such logical operators of weight $8$, but they are all homologically equivalent to each other.  Indeed, not only are they homologically equivalent, but they are equivalent up to a translation symmetry of the lattice (even though the lattice has $16$ sites, an $8$-fold translation of a given representative gives the same operator back, so there are only $8$ representatives at the given weight).  Of course, the same holds for $X$-type logical operators too, since the code is self-dual up to a translation.  So, at weight $8$ there is some logical operator $\tilde Z$ that can be implemented and also some logical operator $\tilde X$ that can be implemented.  We have verified numerically that $[\tilde Z,\tilde X]=0$, i.e., these operators commute.  Hence, \emph{as a stabilizer code, this code has distance $8$, but as a subsystem code, this code has distance $9$.}

This is an interesting phenomenon where the checks all commute with each other (unlike some subsystem codes where the checks do not commute) but we have a larger distance as a subsystem code.  Indeed, there are $4$ logical qubits that are protected to distance $9$ since $\tilde Z$ and $\tilde X$ at weight $8$ commute and so destroy quantum information on a total of $2$ logical qubits.

Another interesting example with  $\sys=4$ is the Hadamard lattice, 
$$
\begin{pmatrix}
1 & 1 & 1 & 1\\
1 & -1 & 1 & -1\\\
1 & 1 & -1 & -1\\
1 & -1 & -1 & 1
\end{pmatrix} \sim 
\begin{pmatrix}
1 & 1 & 1 & 1\\
0 & 2 & 0 & 2\\\
0 & 0 & 2 & 2\\
0 & 0 & 0 & 4
\end{pmatrix}.
$$
The $(2,2)$ is distance is 8, and the code displays a particularly rich set of symmetries and transversal operations which we explore in Sec.~\ref{crystallinegates}.
Unlike the other $\det=16$ lattice above, this code cannot be turned into a distance $9$ code by treating it as a subsystem code; if we
treat all logical operators of weight~$8$ as gauge operators,
then there remain no logical qubits.

For $\sys(L)=5$, an example with optimal determinant is
$$
\begin{pmatrix}
1 & 0 & 1 & 6\\
0 & 1 & 0 & 11\\
0 & 0 & 3 & 9\\
0 & 0 & 0 & 15\\
\end{pmatrix}.
$$
This has $(2,2)$ distance $15$.  We have not tested whether it might have larger distance as a subsystem code, and we have not tested uniqueness of the lattice up to symmetries.

For $\sys(L)=6$, an example with optimal determinant is
$$
\begin{pmatrix}
1 & 0 & 0 & 21\\
0 & 1 & 1 & 24\\
0 & 0 & 2 & 30\\
0 & 0 & 0 & 34\\
\end{pmatrix}.
$$
For $\sys(L)=7$, an example with optimal determinant is
$$
\begin{pmatrix}
1 & 0 & 0 & 115\\
0 & 1 & 0 & 124\\
0 & 0 & 1 & 136\\
0 & 0 & 0 & 152\\
\end{pmatrix}.
$$
We have not tested $(2,2)$ distance or uniqueness of these lattices.

It is interesting to compare these examples to the family of codes defined by matrix
$$\begin{pmatrix} d/2 & d/2 \\ -d/2 & d/2 \\ &&d/2 & d/2 \\ &&-d/2 & d/2\end{pmatrix},$$
where $d$ is even.
This matrix is block diagonal, each block being an optimal matrix for the two-dimensional toric code at given distance.
This family has $\sys=d$ and has $\det=d^4/4$ so $\det/\sys^4=0.25$.
Also, we have $(2,2)$ distance $\leq d^2$ in this case, as it is the homological product of a two-dimensional toric code with itself.
So, we have $\frac{n}{(2,2) \textrm{distance}^2}\geq 1.5$.
So, the examples above lead to some improvement in $\frac{n}{(2,2) \textrm{distance}^2}$ compared to this, and a more significant improvement in $\sys$.

If we consider easily sliceable four-dimensional examples with $\nslice=2$, the optimal $\det$ is shown in \cref{disttableslice4d}.

\begin{table}[h!]
\caption{Easily Sliceable Examples}
\begin{tabular}{l|l}
$\sys(L)$ & $\det(L)$ \\
\hline
3 & 14 \\
4 & 24 \\
5 & 54 \\
6 & 84 \\
7 & 166
\end{tabular}
\label{disttableslice4d}
\end{table}

We can also consider even higher dimensions.  In \cref{higherdim}, we give an explicit family in higher dimensions which are powers of $2$, and we prove optimal scaling of $\sys/\det^{1/D}$ in $D$ dimensions.

\subsection{Starfish syndrome extraction circuit for three-dimensional and four-dimensional cubic lattice}
We now propose an order for measuring $Z$ stabilizers in three-dimensional cubic lattices.  We consider quantum circuits in which a $Z$ stabilizer (which is associated with a plaquette of the lattice) is measured by initializing an ancilla in the state $X=+1$, then applying CZ gates from qubits on each edge of the plaquette to the ancilla qubit, and finally measuring the ancilla qubit in the $X$ basis.  (One may apply a Hadamard to the ancilla and instead this becomes the more usual circuit in which the ancilla is prepared in the $Z=+1$ state and we use CNOT gates from data qubits to the ancilla; we use this notation simply so that all qubits are initialized in the $X=+1$ state for notational convenience.)

We have to pick an order to apply the CZ gates.
The importance of choosing this order is that a single $X$ error on the ancilla qubit can become several $Z$ errors on the data qubits.

We propose to use the following order that we call starfish.  There are three directions for the three-dimensional lattice, which we call directions $1,2,3$.  We imagine that each ancilla qubit is located in the center of the corresponding plaquette and we imagine that each data qubit is located in the center of the corresponding edge; this need not be the physical locations of the qubit, but we use these locations to describe the order as then each ancilla is separated from the four data qubits in the corresponding stabilizer by a vector pointing in one of the three lattice directions.
We first apply CZ gates from data qubits to ancilla such that the ancilla is displaced from that data qubit in the positive $1$ direction, then in the negative $1$ direction, then in the positive $2$ direction, then in the negative $2$ direction, then in the positive $3$ direction, and finally in the negative $3$ direction. 
We will denote this gate ordering with the shorthand $(+1,-1,+2,-2,+3,-3)$. 
Of course, any given ancilla qubit will only participate in four of these six rounds.

An $X$ error on an ancilla may propagate to put a $Z$ error on the data qubits.
If this ancilla $X$ error occurs after exactly two CZ gates are done,
then the propagated error has weight two acting on the east and west data qubits.
The syndrome by this propagated weight-$2$ error is four-vertex violations.  This is the reason for choosing this order, as it maximizes the number of violations for this kind of error.  If the ancilla qubit $X$ error occurs after one of three CZ gates then it causes either a weight-$1$ error (which is fine as it does not increase the error weight) or a weight-$3$ error (which is equivalent to a weight-$1$ error up to a stabilizer).

We have verified that this order preserves the distance (i.e., that the distance with circuit level noise is equal to the $\ell_1$ systole) for distance $3$ and $5$ codes.

We can also use the starfish ordering in the four-dimensional (2,2) toric code, we can measure the stabilizers sequentially by applying all two qubit gates for the X-stabilizers in the directions $(+1,-1,+2,-2,+3,-3,+4,-4)$, and similarly for the Z-stabilizers. 
This results in a syndrome extraction circuit of depth~$16$
($8$ for $X$-stabilizers and $8$ for $Z$-stabilizers)
for the four-dimensional $(2,2)$ toric code.


\section{A method for logical bell pair creation}
\label{pathtocliffords}

Here we present a method for creating a logical Bell pair.
Once we can create logical Bell pairs, then this can be used for error correction\cite{knill2007quantum}.

If our initial code is a $(2,2)$ toric code in four dimensions (or an even higher dimensional toric code), then a fixed logical state (e.g., the $+$ state) can be created fault tolerantly in a single shot manner.  For a three-dimensional code, however, it is only possible to have single-shot fault tolerance against one kind of error.  However, by preparing the data qubits in the $X=+1$ product state, and then measuring $Z$ stabilizers on plaquettes (which are single-shot fault tolerant), we can prepare the code state in a single shot manner.

We will explore primarily this three dimensional case in this section.

We introduce two main ideas in this section.  The first idea is that of slicing: we first create a three-dimensional toric code state on the 3-torus, and
we then ``slice" open this 3-torus into two code states which are entangled in a Bell pair (or several bell pairs).  This slicing idea generalizes to many other codes.
The second idea is a treatment of fault-tolerant Bell correlation measurement with a referee. This is conceptual rather than practical and is a theoretical setup by which we generate a Bell state and verify it. The issue here is that the generation of the Bell state in our codes requires some global communication in the error correction. We have error chains propagating outside the lower dimensional code blocks and these error chains can be dangerous if we decode the lower dimensional codes completely separately. Then, it could become questionable if our Bell correlation measurement at the logical level was due to correlation introduced in the error correction and if we measured something inherent at the logical state. By introducing a referee in addition to lower dimensional code blocks, we show that Bell inequality violation is possible while obeying the no-signaling rule. This clarifies that our protocol generates a Bell pair that can be used to teleport an arbitrary logical state fault tolerantly without resorting to ``correlated decoding’’~\cite{cain2024correlated}.

\subsection{Protocol}
Our basic protocol to create the logical Bell pairs is as follows:
\begin{enumerate}
    \item Initialize all physical qubits in the $X=+1$ state. Using an integral lattice boundary conditions,
    the total number of qubits is $\abs{3\det M}$ where the rows of~$M$ are a basis for the integral lattice.
    Each qubit is associated with an edge.
    \item Measure all $Z$-stabilizers (each defined on a square) once.
    This may be performed using ancilla qubits and unitary gates.
    \item Apply a correction operator, some tensor product of single qubit Pauli~$X$,
    to make all $Z$-stabilizers assume eigenvalue~$+1$.
    \item Choose two disjoint subsets of qubits, each of which forms a closed 2-torus, 
    such that a representative of a fixed $Z$-logical operator is supported on each subset,
    i.e., the product of the two representatives is a stabilizer of the three-dimensional toric code.
    \item Measure all qubits in~$X$ that are not in either of the two subsets.
    \item Up to a Pauli correction, each subset holds a toric code state, jointly in a Bell state. 
\end{enumerate}

In \cref{slicing} we analyze the correlation of the logical qubits in the absence of error.  
In \cref{errorcorrect}, we analyze error correction.
An interesting distance becomes relevant for error correction.
When we measure the $Z$ stabilizers, the outcomes are random.
However, in the absence of errors, the $Z$ stabilizer measurement outcomes form a homologically trivial $2$-cocyle (flux loops).
Hence,
these will be detected unless one has at least $2d$ mismeasured stabilizers.
The factor of~$2$ in~$2d$ arises as follows: a total of $d$ measurement errors can form a $2$-cocycle which ``wraps" around the torus once, as the $\ell_1$ systole is equal to $d$.  However, such a cocycle would not be homologically trivial, so one needs $2d$ mismeasured stabilizers to have an undetectable error.

\subsection{Slicing}
\label{slicing}

To find the bipartition, it is simplest if we have an easily sliceable example with $\nslice=2$.

If we have an easily sliceable example, with $\nslice=2$, and if we measure all qubits corresponding to edges in the $1$ direction, measuring in the $X$-basis, then (in the absence of errors) the resulting logical state consists of two Bell pairs, with each of the two logical qubits of the two-dimensional toric codes in a Bell pair with a logical qubit of the other two-dimensional toric code.

To see this, let $\tilde Z_1$ be a $Z$-type logical operator in the first two-dimensional toric code corresponding to a cycle in some direction of the torus, and let $\tilde Z_2$ be the corresponding $Z$-type operator in the second two-dimensional toric code.  Then, the product $\tilde Z_1 \tilde Z_2$ is a homologically trivial cycle \emph{in the three-dimensional toric code}, and hence has expectation value $+1$.  Similarly, let $\tilde X_1$ be an $X$-type logical operator in the first two-dimensional toric code and let $\tilde X_2$ be the corresponding operator in the second two-dimensional toric code.  Then, the product $\tilde X_1 \tilde X_2$ is a logical operator of the three-dimensional toric code and hence also has expectation value $+1$.

We can easily extend this calculation to $\nslice>2$.  In this case, one finds that the logical qubits are in a GHZ state.  Let $\tilde Z_i^a$, for $i\in \{1,\ldots,\nslice\}$ and $a\in {x,y}$ be a $Z$ type logical operator on the $i$-th two-dimensional toric code corresponding to a cycle in either the $x$ or $y$ direction of the lattice.  Let $\tilde X_i^a$ be $X$-type logical operators, with $[\tilde Z_i^a,\tilde X_i^a]=0$ so that $\tilde X_i^a$ is a cocycle in the other direction of the two-dimensional lattice.  Then, the logical qubits are stabilizers by $\tilde Z_i^a \tilde Z_j^a$ for all~$i,j,a$ and 
by~$\prod_i \tilde X_i^a$ for all $a$.

This calculation is a general case of the calculation in \cref{slicetwist} which explains this for more general codes.

\subsection{Error correction}
\label{errorcorrect}
We now explain how error correction must work to verify the logical Bell correlation.

\paragraph{Logical $X$ measurement---}

Since we aim to measure the $X$ logical operator on both two-dimensional toric codes,
we measure every qubit in~$X$ on the prepared three-dimensional toric code ---
the slicing is implemented by single-qubit $X$ measurements in the bulk
and the logical $X$ measurement is by single-qubit $X$ measurements on two-dimensional toric codes.
Any fault configuration is divided into two categories.
One consists of measurement outcome flips of $Z$-stabilizers on squares (flux errors)
and the other of single-qubit $Z$ errors on qubits that comprise the three-dimensional toric code.
The single-qubit $X$ measurements may be faulty,
but these errors can be interpreted as if there a $Z$ error occurred
in qubits that are measured.

A flux error is benign for this logical $X$ measurement
because the zero flux configuration will be realized by applying certain $X$ operator,
which will be completely ignored upon single-qubit $X$ measurements.
So we only have to worry about single-qubit $Z$ errors,
and we are effectively given perfect measurement of all $X$-stabilizers of the three-dimensional toric code.

This is a well studied setting for error correction:
a configuration of error defines a 1-chain in the cubic lattice
with periodic structure (boundary conditions) defined by our integral lattice,
and nontrivial logical errors occur only if the decoding results in a homologically nontrivial 1-cycle,
that intersects 1-cohomology class of one of the two-dimensional toric code slices.
The usual matching decoding algorithm among other methods can be used.

\paragraph{Logical $Z$ measurement---}

Now what matters is $X$ errors on qubits and flux errors.
We discuss them in order.

In the bulk, all $X$ errors are absorbed to the single-qubit measurements, so we should worry about $X$ errors on the data qubits of the two-dimensional toric code slices. Since all qubits of the two-dimensional toric code are measured in the $Z$ basis, we infer the flux values on every plaquette reliably, and these are our syndrome bits. Any undetectable errors must come from some (dual) $X$ string logical operator on the two-dimensional slice. Such an error may appear by qubit $X$ errors on the two-dimensional slice, which requires d errors. Usual matching algorithm or any other two-dimensional toric code decoding algorithm can be used to achieve fault tolerance.
 
Flux errors (incorrect $Z$-plaquette measurements) indirectly introduce $X$ errors. When we measure the flux values before slicing, the requirement that flux must form a zero-homology cycle allows us to identify some flux errors and correct them. The correction will only be correct up to a zero-homology cycle that bounds an $X$ chain. This $X$ chain is not visible at the 3d code preparation stage but can be detected when we destructively measure the two-dimensional toric code slice in the $Z$ basis. For flux error to go undetected at the 3d code preparation stage and leave an $X$ error, at least two flux errors are needed on both sides of the two-dimensional toric code slice, orthogonal to the slice. Hence, $e_f$ flux errors and $e_X$ qubit $X$ errors result in a logical undetected error only if $\frac 1 2 e_f + e_X \ge d$. The coefficient $\frac 1 2$ is sharp here.

\begin{table}[b]
\begin{tabular}{c|cc}
effective code distance &  $5$ & $7$ \\
\hline
Hermite normal form  &
    $ \begin{pmatrix}
        2 & 0 & 12  \\
        & 1 & 8 \\
        & & 13
    \end{pmatrix}$
    &
    $ \begin{pmatrix}
        2 & 0 & 12  \\
        & 1 & 7 \\
        & & 25
    \end{pmatrix}$
\end{tabular}
\caption{Minimal examples for odd effective code distances for Bell pair generation}
\label{tb:sliceable-shallow}
\end{table}

\paragraph{Reducing qubit count---}

We have determined that the effective code distance for qubit errors~$Z$ 
is given by the minimal homology $1$-cycle that intersects a $1$-cohomology that is transverse to the two-dimensional toric code,
and the effective code distance for qubit error~$X$ is given by
the minimal $1$-cohomology representative of the two-dimensional toric code,
while the flux errors (flips in measurement outcomes of $Z$-stabilizers on squares) have effective code distance
twice the $\ell_1$ systole.
This allows for further optimization of the qubit count given a target effective code distance.
\Cref{tb:sliceable-shallow} contains optimized examples for effective code distances~$5$ and~$7$.
The easily sliceable instance with~$d=3$ is already optimal.

\paragraph{Toward Bell's inequality violation---}

To this end, some modification is necessary
because the error correction explained above requires aggregate of all measurement data
which include all logical information.
The following modification in error correction enables fault tolerance 
in a hypothetical scenario where two two-dimensional toric codes are separated far away from each other,
and can demonstrate Bell's inequality violation using logical qubits.
We imagine that there are three parties involved.
Two parties $A$ and $B$ will hold two logical qubits that are in a Bell state
and measure correlation.
The third party~$C$ is a referee that prepares a logical Bell state,
but do not measure Bell correlation at the logical level.

The referee prepares all individual qubits in~$X=+1$ state,
and measure square plaquettes in~$Z$.
The referee applies $X$-corrections to ensure that 
the resulting state is the three-dimensional toric code state with logical $\tilde X = +1$.
Then, $C$ measures out the bulk qubits in $X$ to generate the two two-dimensional toric code states.
Let us say that one toric code is $A$'s and the other is $B$'s.
The referee~$C$ sends out these data qubits to~$A$ and~$B$, respectively.
In addition to the data qubits of $A$'s and $B$'s codes,
$C$ sends the single-qubit measurement outcomes near~$A$ to~$A$ and those near~$B$ to~$B$.
For large code distances~$d$, qubits within distance, say,~$d/5$ from $A$ and $B$ are deemed near~$A$ and $B$, respectively.
The referee~$C$ runs its own decoding algorithm to correct against $Z$ errors that are likely supported away from~$A$ and~$B$ by distance, say,~$d/10$.
Parties~$A$ or~$B$ now perform their logical measurement fault-tolerantly using their own measurement outcomes and the outcomes that $C$ has sent them.
Players $A$ and $B$ report their logical outcome to $C$, who combines the results and confirms Bell correlation.

In this scenario, the referee cannot infer the measurement outcome of either~$A$ or~$B$ beforehand,
and $A$ and $B$ do not have access to the other party's qubits.
Under a local stochastic error model, it is unlikely with large $d$ that the decoding will fail
because usually there will not be any long error chain that traverses the boundary regions of the window decoding.

\section{Logical Gates from Crystalline symmetry}
\label{crystallinegates}
In this section we describe how to generate logical operations using crystalline symmetries.
We describe a set of code automorphisms which are found by putting together the results of~\cite{breuckmann2024fold} and the use of crystalline symmetries. 
We also discuss a new type of lattice surgery, enabling more logical operations. 

\subsection{Notation}

In what follows it is helpful to use the symplectic representation of the Pauli and Clifford group.
The isomorphism between the Pauli group $P_n$ on $n$ qubits modulo phases denoted $\hat{P}_n = P_n/\langle 1, i,-i,-1 \rangle$, and the symplectic vector space $\left(\mathbb{F}_2^{2n},J \right)$ is given by 
\begin{align}
(\text{phase}) X({\bf a}) Z({\bf b}) \mapsto ({\bf a}|{\bf b}),
\end{align}
where $X({\bf a}) = \bigotimes_{j = 1}^{n} X_j^{a_j}$, $Z({\bf b}) = \bigotimes_{j = 1}^{n} Z_j^{b_j}$, and $({\bf a} | {\bf b}) = {\bf a} J {\bf b}^T$.
Multiplication in the group $\hat{P}_n$ becomes addition in the symplectic vector space.
The commutation relations of the Pauli matrices are encoded in the symplectic form 
\begin{align}
    J_n = \begin{pmatrix}{\bf 0}&\text{id}_n\\ \text{id}_n &{\bf 0} \end{pmatrix}.
\end{align}
If $P$ and $P'$ are two Pauli operators, $v$ and $v'$ the corresponding $\mathbb{F}_2$ valued vectors, and $PP' = (-1)^{\text{comm}(P,P')} P'P$ then $\text{comm}(P,P') = v J v'^T$.
In what follows, we will leave $n$ implicit.
Note that we are representing the elements of the Pauli group by a $2n$ dimensional row vector.

A stabilizer code $\mathcal{C}$ is specified by an Abelian subgroup of the Pauli group which does not include $-\text{id}$. 
In the symplectic representation, the code is specified by a $m$ by $2n$ parity check matrix $C = (C_x|C_z)$ satisfying $CJC^T = {\bf 0}$.
A CSS code is any code which can be written as, 
\begin{align}
C = 
\begin{pmatrix}
    \begin{array}{c|c}
        C_X & 0 \\
        0 & C_Z \\
    \end{array}
\end{pmatrix}.
\end{align}

A logical operator ${\bf v} \in \mathbb{F}_2^{2n}$ is any operator satisfying $C J {\bf v}^T ={\bf 0}$.
The number of logical qubits is given by $k = n - \text{rank} C$.
It is convenient to package a generating set of logical operators into a $k$ by $2n$ matrix $L$ satisfying $LJL^T = J$ and $CJL^T = 0$. 
Such a basis always exists.
It is convenient to think of the matrix $L$ as a map from the Pauli group on $k$ qubits to the code space of $C$.

A symmetry of the code is given by a unitary matrix $U$ which takes the code space back to itself. 
Here, we consider the subset of symmetries given by unitary Clifford matrices. 
A Clifford unitary is any unitary matrix which maps products of Pauli operators to products of Pauli operators.
It is again convenient to work in the binary symplectic representation, where we consider Clifford unitaries modulo the group $\hat{P}_n$.
We will still refer to this as the clifford group, with the implicit understanding that we are working modulo the Pauli group mod phases. 
Correspondingly, a Clifford unitary is represented by a $2n$-by-$2n$ symplectic matrix. 
That is a $2n$-by-$2n$ matrix $U$ is Clifford if $U J U^T = J$, i.e., the group of Clifford matrices modulo $\hat{P}_n$ is the symplectic group
\begin{align}
\text{Sp}(2n,\mathbb{F}_2) = \left \{ U \in \mathbb{F}_2^{2n \times 2n} : UJU^{T} = J \right \}
\end{align}
A Clifford symmetry of the code is any Clifford unitary $U$ such that \begin{align}
\text{Aut}(C) = \left \{ U \in \text{Sp}(2n,\mathbb{F}_2):  \text{RowSpan}(C U) = \text{RowSpan}( C) \right \}.
\end{align}

\subsection{Review of permutation and fold-transversal symmetries}
In this subsection we review three types of symmetries: permutation, and the fold transversal symmetries of Hadamard- and Phase-type\cite{breuckmann2024fold}. 

An $n$ by $n$ permutation matrix $P$ acts on the qubits via the matrix,
\begin{align}
U(P) =\begin{pmatrix}P& 0\\ 0 &P \end{pmatrix}
\end{align}
The qubit permutation $P$ is a symmetry of the code if the row span of $C$ and $CU$ are identical. 

For a CSS code built from parity check matrices $C_X$ and $C_Z$, a ZX duality~\cite{breuckmann2024fold} is any permutation $D$ of the qubits which maps $\text{RowSpan}(C_X)$ surjectively into $\text{RowSpan}(C_Z)$ and vice versa. 
That is $\text{RowSpan}(C_X D)= \text{RowSpan}(C_Z)$ and $\text{RowSpan}(C_Z D)= \text{RowSpan}(C_X)$.

Given a ZX duality $D$ one can define the Hadmard-type symmetry, 
\begin{align}
\label{eq:foldH}
H_D=
\begin{pmatrix} 
0&D\\
D&0\\
\end{pmatrix} 
\end{align}
To check that $H_D$ is a symmetry we note that $C H_D = \left(\begin{smallmatrix}0& C_X D\\ C_Z D &0 \end{smallmatrix}\right)$ which shares the same row span as $C$ since $\text{RowSpan}(C_Z D)= \text{RowSpan}(C_X)$ and vice versa. 
In words, we find a qubit permutation which maps the support of each X-stabilizer to the support of a Z-stabilizer, apply the permutation to the qubits, followed by a transversal Hadamard gate.

Similarly, given a ZX duality $D$ such that $D^2 = \text{id}$, one can define the Phase-type gate $S_D$  as,
\begin{align}
\label{eq:foldS}
S_D = 
\begin{pmatrix} 
\text{id}&D\\
0& \text{id}\\
\end{pmatrix}.
\end{align}
In words, we find a qubit permutation which interchanges the support of each $X$-stabilizer to the support of a $Z$-stabilizer, and then apply a $CZ$ to any qubits which are identified under this map, followed by an $S$ gate on all qubits which are left invariant.
We remark that the constraint $D^2 = \text{id}$ can be relaxed as long as $\text{RowSpan}(C_X D) \subseteq \text{RowSpan}(C_Z)$.

\subsection{Symmetry action on the logical subspace}

Given a logical subspace $L$ of a code $C$, we can compute the logical action directly by applying one of the above unitaries, and computing how the transformed logicals commute with the initial logical operators. 
Under the unitary $U$, we note that $L \mapsto LU$, and since $U$ is a symmetry we have $\text{RowSpan}(LU) = \text{RowSpan}(L)$, and therefore there exists an $M$ such that $LU = ML$.
Using that $LJL^T=J$ we have $LUJL^T = MJ$ and so, 
\begin{align}
M = LUJL^TJ. 
\end{align}
Where $M$ is the action of $U$ on the logical operators. 
Note that $M \in \text{Sp}(2k,\mathbb{F}_2)$ is a $2k$ by $2k$ matrix which acts from the right on row vectors of the logical Pauli group.

\subsection{Logical operations from crystalline symmetries}

Computing the permutation, Hadmaard-type and Phase type symmetries is difficult to do by brute force when the code size becomes too large. 
For this reason we resort to spatial symmetries. 
Of course, not all codes admit spatial symmetries, but when they do this gives a powerful method for computing logical operations.
For simplicity, in the following we will restrict our attention to translation invariant codes on the $d$ dimensional torus, but the discussion can be straightforwardly extended to $d$ dimensional cubes or any code which can be brought to a form that admits an action of a space group.

To setup the problem, we consider a translation invariant code $C$ in $d$ dimensions with periodic boundary conditions.
We specify the periodic boundary conditions by a $d$-dimensional integer matrix $W$ of rank $d$.
The $d$ dimensional torus is given by the quotient $\mathbb{Z}^d/W$.
That is, two points in the $d$-torus are equivalent if they are related by an integer combination of row vectors of $W$.
Note that if $U$ is a unimodular matrix (integer matrix with determinant equal to $+1$ or $-1$), then $UW$ is an equivalent lattice, and correspondingly describes an equivalent periodicity constraint, and an equivalent code.

Since the code is translation invariant, we have some number of qubits in each unit cell. 
It is convenient to assign each qubit in a unit cell a coordinate ${\bf r}_j$ for $j =1, \cdots, q$.
We have no constraints on the ${\bf r}_j$ other than those imposed by translation invariance (e.g., ${\bf r}_j$ may be equal to ${\bf r}_k$ even when $k\neq j$). 
Similarly we will assign each stabilizer a coordinate as well which respects the unit cell.
This coordinate assignment is natural for many topological codes, as in many cases they can be placed on a translation invariant cellulation which naturally provides coordinate positions. 
We will give concrete examples in the following subsections. 
Having situated each qubit at a physical coordinate along with translation invariance,
we have a ``crystal of qubits''. 
As we will see, we can search for symmetry operations by exploiting the crystalline geometry.

Recall in $d$ dimensions the space group is the set of rigid transformations which leave a translation invariant set of points invariant, it is a subgroup of the $n$-dimensional Euclidean group $E(n)$.
A space group transformation is labeled by an orthogonal rotation $M \in O(d,\mathbb{R})$ along with a shift $b \in \mathbb{R}^d$. 
Its action on a vector is given by: 
\begin{align}
(M, {\bf b}): {\bf r} \mapsto {\bf r}M + {\bf b}.
\end{align}
Where we have taken the vector ${\bf r}$ to be a row vector. This is related to the fact that our integer matrix $W$ has row vectors as lattice vectors.
The group law is given by, 
\begin{align}
(M',{\bf b}') \circ (M,{\bf b}) = (MM',{\bf b}M'+{\bf b}').
\end{align}
The subgroup of the space group formed by elements of the form $(M,{\bf 0})$ is called the point group. 
The space group itself consists of all transformations of the form $(M,{\bf b})$ which preserve the set of of crystalline coordinates. 

We now consider how we can act on a stabilizer group with a crystalline symmetry. 
First, the space group symmetry must be compatible with the periodicity constraints imposed by $W$.
This imposes the constraint that the lattice generated by $W$ is preserved by the orthogonal matrix appearing in the space group transformation.
That is, we require 
\begin{align}
\text{hnf}(WM) = \text{hnf}(W).
\end{align}
Where $\text{hnf}(\cdots)$ takes the Hermite normal form of the matrix it acts on. 
An equivalent condition is requiring the existence of a unimodular matrix $U$ such that $WM = UW$.
We only consider the set of space group elements $(M,b)$ which satisfy $\text{hnf}(WM) = \text{hnf}(W)$.
Such an $M$ is called a lattice automorphism.

In low dimensions the space groups have been classified.
In general dimensions the space group is not classified, hence we provide a simple algorithm for determining elements of the space group below.

\subsubsection{Algorithm for finding space group symmmetries}
To determine a set of space group transformations we use the following algorithm. 
We begin by computing the group of lattice automorphisms.
Since these automorphisms must preserve both the lengths of the basis vectors and their inner products, each basis vector has only finitely many possible image vectors.  
This allows for an exhaustive search in low dimensions, though the complexity remains exponential in the lattice dimension.
In practice, we use SageMath to directly compute group of lattice automorphisms\cite{sagemath2023}.
We consider a unit cell, and the set of coordinates ${\bf r}_j$ with $j = 1, \cdots, q$. 
Pick a coordinate, say ${\bf r}_1$, and a lattice automorphism $M$.
We know that ${\bf r}_1$ must be mapped to some ${\bf r}_j + \Delta$ under a space group transformation $(M,{\bf b})$, where $\Delta$ is an integral vector.
To determine this set, we simply pick a pair of points ${\bf r}_j$ and ${\bf r}_k$ and solve for ${\bf b}$, and then check whether it satisfies all the constraints of a space group symmetry.
Since this is a finite problem, it is efficient to solve on a computer. 
The resulting $q^2$ linear equations are straightforward to solve for a given $M$:
\begin{align}
(M,{\bf b}_{kj})({\bf r}_j) = {\bf r}_k \quad \implies \quad {\bf b}_{kj} = {\bf r}_k - {\bf r}_j M.
\end{align}
Each candidate pair $(M,{\bf b}_{kj})$ can then be checked explicitly if it is a space group symmetry. This solution fixes a particular $\Delta$, but any other choice of $\Delta$ is related by an integral vector, and result in an equivalent space group transformation with a shifted origin.

\subsubsection{Crystalline permutation symmetry}
A crystalline permutation symmetry is given by the space group action on the qubits which preserves the stabilizer group.
A generalized crystalline permutation symmetry is given by $(M,{\bf b},U)$ where $U$ is a finite depth local Clifford circuit, and $(M,{\bf b})$ is a space group transformation. 
Explicitly, it is given by, 
\begin{align}
    (M,{\bf b},U): P_{{\bf r}_j} \mapsto U
    P_{{\bf r}_j M+{\bf b}}
    U^{\dagger}.
\end{align}
Where $P \in \{ \text{id}, X, Y, Z \}$. 
Of course, setting $U = \text{id}$, $M = \text{id}$, and ${\bf b}$ to any coordinate direction will be a symmetry due to translation invariance.
When $U$ is trivial, this is purely a permutation symmetry.

\subsubsection{Crystalline Hadamard-type symmetry}
Given a CSS code we can define a Hadamard-type symmetry using the crystalline space group. 
Recall that a Hadamard-type symmetry uses a ZX duality as input. 
A ZX duality can be found by looking for the set of space group transformations $(M,{\bf b})$ which map X-stabilizers to Z-stabilizers, Z-stabilizers to X-stabilizers, and qubits to qubits.
Once one finds the set of space group transformations satisfying these constraints, one can construct the Hadamard-type symmetry as shown in Eq.~\ref{eq:foldH}.

\subsubsection{Crystalline Phase-type symmetry}
Given a CSS code we can define a Phase-type symmetry using the crystalline space group. 
We simply search for the set of space group elements $(M,{\bf b})$ which map X-stabilizers to Z-stabilizers, and map qubit locations to the same set of qubit locations.
We also require that the map is order $2$, i.e., $M^2 = \text{id}$ and $M{\bf b}+{\bf b} = {\bf 0}$. 
Once one finds this subset of ZX dualities, one can construct the Phase-type symmetry following Eq.~\ref{eq:foldS}.

\subsection{Examples}
\subsubsection{The two-dimensional rotated toric code}
\label{clifford2d}
First, we can realize $2\nslice$ logical qubits by taking $\nslice$ slices of a three-dimensional toric code on the three-torus. 
In this subsection we describe how to apply a few examples of fold-Clifford circuits to these $2\nslice$ logical qubits.

Remark: in certain instances we may prefer to compile $\nslice$ logical qubits into $\nslice$ slices of the 3 torus.

The first example we look at is given by slicing into the two-dimensional toric code described by 
\begin{align}
\begin{pmatrix}
    n&0\\ 0&n
\end{pmatrix}.
\end{align}
A Hadamard-type gate is given by the qubit permutation found by translating all the qubits by the vector $(\pm 1/2, \pm 1/2)$, and applying the transversal Hadmard gate.  
A Phase-type gate can be found with a ZX duality given by a reflection across the line $\alpha (1/2,1/2) + (1/2,0)$, with $\alpha $ real.
In this case the group of logical operations generated by crystalline symmetries is given by $H \otimes H$, $\text{SWAP}_{12}$, $S \otimes S$, and $CZ_{12}$.

Examples of three-dimensional toric codes which slice into two-dimensional toric codes with the above gate set have Hermite normal form, 
\begin{align}
\begin{pmatrix}
    2&1&1\\ 0&3&0\\ 0&0&3
\end{pmatrix},
\quad
\begin{pmatrix}
    2&1&1\\ 0&5&0\\ 0&0&5
\end{pmatrix},
\quad 
\begin{pmatrix}
    2&5&3\\ 0&7&0\\ 0&0&7
\end{pmatrix}
\end{align}
describing codes with parameters [[54,3,3]], [[150,3,5]], and [[294,3,7]].
The codes are CSS and so logical CNOT is transversal.

\subsubsection{The four-dimensional (2,2) toric code}
We consider the four-dimensional (2,2) toric on a rotated cubic lattice defined by $W$. 
The rotated cubic lattice has 0,1-,2-,3-,and 4-cells.
We can assemble these cells into the following 5-term chain complex, 
\begin{equation}
\begin{tikzcd}
C_4 \arrow[r, bend left=20, "\partial_4"] & 
C_3 
\arrow[l, bend left=20, "\partial^T_4"]
\arrow[r, bend left=20, "\partial_3"] &
C_2
\arrow[l, bend left=20, "\partial^T_3"]
\arrow[r, bend left=20, "\partial_2"] &
C_1 
\arrow[l, bend left=20, "\partial^T_2"]
\arrow[r, bend left=20, "\partial_1"]&
C_0
\arrow[l, bend left=20, "\partial^T_1"]
\end{tikzcd}
\end{equation}
Where $\partial$ is the boundary operator and satisfies $\partial^2 =0$, and $\partial^T$ is the coboundary operator composed with the isomorphism between $C_j$ and $C^j$, as a matrix it is the transpose of $\partial$.
Any consecutive triple of elements in the chain complex defines a CSS code. 
It is convenient to label each j-cell, by a point $p$, along with $j$ integers in the range $\{0,1,2,\cdots, d-1\}$. 
These integers represent the cells of interest, for example $(p,\{0,1\})$ denotes the two-cell which contains the point $p$, and the edges $p+\hat{0}$ and $p+\hat{1}$.
Where $\hat{i}$ is $i$th positively oriented unit vector in $\mathbb{R}^4$.

The (2,2) toric code uses $C_3,C_2$, and $C_1$, placing qubits on the 2-cells, X-stabilizers on the 1-cells, and Z-stabilizers on the 3-cells. 
For a 1-cell ${\bf c}_1 \in C_1$ the X-stabilizer is given by $X(\partial^T_2 {\bf c}_1)$.
And we have used $X({\bf v}) = \bigotimes_{j} X_j^{v_j}$.
Similarly, for a 3-cell ${\bf c}_3 \in C_3$ the Z-stabilizer is given by $Z(\partial c_3)$.
This is a CSS code, and the parity check matrix is simply, 
\begin{align}
C = 
\begin{pmatrix}
    \begin{array}{c|c}
        C_{\partial^T_2} & 0 \\
        0 & C_{\partial_3} \\
    \end{array}
\end{pmatrix}.
\end{align}
where $C_{\partial^*}: C_1 \rightarrow C_2$ and $C_\partial: C_3 \rightarrow C_2$ are the matrices representing the boundary and dual boundary operators respectively.  

Now lets define the crystal. We label each stabilizer and qubit by the midpoint of its corresponding cell.
That is: 
\begin{align}
(p,\{i,j,k\}) &\mapsto p+\frac{\hat{i} +\hat{j} + \hat{k}}{2}\\
(p,\{i,j\}) & \mapsto p+\frac{\hat{i}+ \hat{j}}{2}\\ 
(p,\{i\}) &\mapsto p+\frac{\hat{i}}{2}
\end{align}
where $p \in \mathbb{Z}^4/W$.

There are many interesting explicit examples to consider, a particularly interesting one is the Hadamard lattice given by, 
\begin{align}
\label{eq:hadamardLattice}
W = 
\left(
\begin{array}{cccc}
1 & 1 & 1 & 1 \\
1 & -1 & 1 & -1 \\
1 & 1 & -1 & -1 \\
1 & -1 & -1 & 1
\end{array}
\right)
\sim 
\left(
\begin{array}{cccc}
 1 & 1 & 1 & 1 \\
 0 & 2 & 0 & 2 \\
 0 & 0 & 2 & 2 \\
 0 & 0 & -2 & 2 \\
\end{array}
\right)
\sim 
\left(
\begin{array}{cccc}
 1 & 1 & 1 & 1 \\
 0 & 2 & 0 & 2 \\
 0 & 0 & 2 & 2 \\
 0 & 0 & 0 & 4 \\
\end{array}
\right).
\end{align}
On the right we have put $W$ into its Hermite normal form. 
The corresponding four-dimensional (2,2) toric code is a block code $[[96, 6,8]]$ with 64 X-stabilizer and 64 Z-stabilizers.

\begin{figure}
    \centering
\includegraphics[width=0.95\linewidth]{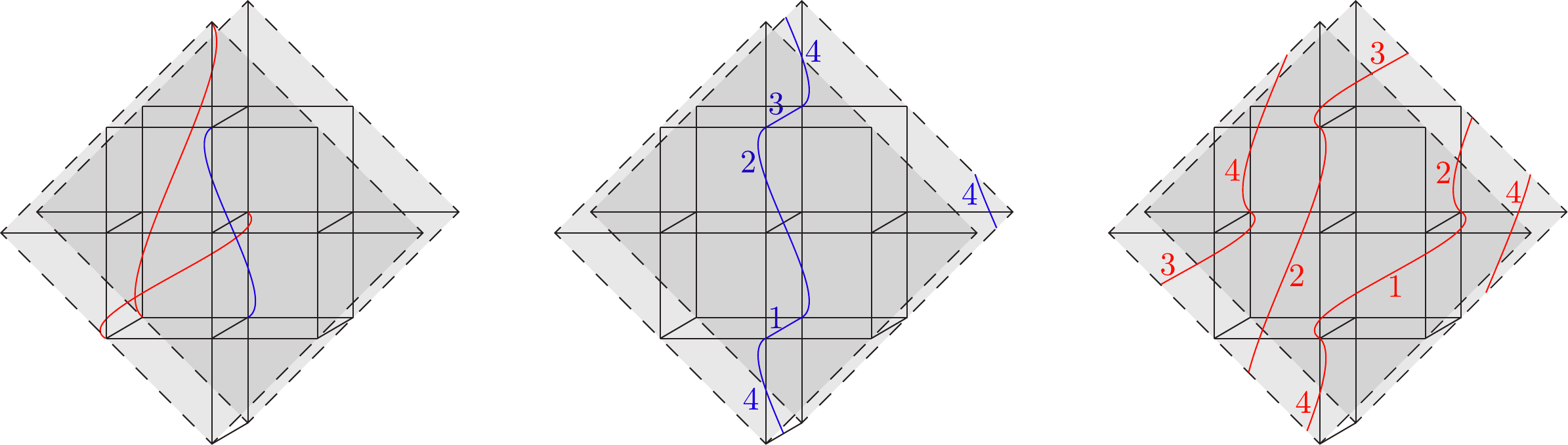}
    \caption{A diagram of the connectivity of the 1-cells in the Hadamard lattice. 
    A convenient basis is shown in the middle of Eq.~\ref{eq:hadamardLattice}.
    The bottom $2$ by $2$ matrix is represented by the diamond shape, and the corresponding connectivity---it is the lattice used for a rotated two-dimensional toric code.
    Note that the upper left boundary is connected to the lower right, and the lower left boundary is connected to the upper right boundary.
    There are 8 vertices associated with it. 
    The connectivity in the 3rd and 4th dimension is twisted.
    The $2$ on the diagonal in the second row the integral matrix means we have ``two layers'', hence why we have drawn two overlaid diamonds. The two layers have a direct connection as drawn by the black lines connecting the layers and also a twisted boundary condition shown with a blue. 
    The last $2$ in the second row means when we translate from the back diamond to the front we shift  $2$ along the vertical direction.
    We have drawn the corresponding edge in blue, all others are found by translates of this. 
    The first row $(1,1,1,1)$ means that there is an edge in the fourth direction but shifting in the $(1,1,1)$ direction once reconnecting.
    Being mindful of the twisted boundary conditions, we have drawn two representative edges in red. 
    The Hadamard lattice is especially symmetric, the four-dimensional analogue of the rotated two-dimensional toric code. In particular any straight line following the 1-cells has $\ell_1$ distance 4. We have drawn two non-trivial 1-cycles on the right in blue and red. 
Any horizontal or vertical line in the gray diamond will also have $\ell_1$ distance 4.
    }
    \label{fig:hadamardlattice}
\end{figure}

This code is nice because it is particularly symmetric.
The lattice $W$ has $384$ lattice automorphisms.
The lattice automorphisms yield permutation symmetries, of which only $24$ are distinct.
In Fig.~\ref{fig:hadamardlattice} we have shown the connectivity of the 1-cells of the lattice using the central matrix in Eq.~\ref{eq:hadamardLattice}. 

The Hadamard-type gates are more interesting. 
The four-dimensional cubic lattice has a Poincare duality. 
This duality maps $\star: C_j \rightarrow C_{d-j}$.
In general, and using the notation for an $l$-cell the Poincare duality acts on the crystal as, 
\begin{center}
\begin{tikzcd}[row sep=2.5em, column sep=2.5em]
(p,\{i_1,\cdots, i_l\}) \arrow[r, "\star"] \arrow[d, "\text{pt}"'] & (p+\hat{i}_1 + \cdots + \hat{i}_l,\{i_1,\cdots, i_l\}^*) \arrow[d, "\text{pt}"] \\
p + \frac{\hat{i}_1 + \cdots + \hat{i}_l}{2} \arrow[r, "\star"'] & p+\hat{i}_1 + \cdots + \hat{i}_l + \frac{1}{2} \sum_{\hat{j} \notin \{\hat{i}_1, \cdots , \hat{i}_l  \}} \hat{j}
\end{tikzcd}
\end{center}
where the $\text{pt}$ function takes the midpoint of the $l$-cell.
And $*$ means to take every coordinate direction not in the tuple.
Up to qubit permutation symmetries, the Hadamard-type gate coming from Poincare duality is the only Hadmard-type gate found, up to qubit permutation.  
The space group transformation and corresponding logical action given by 
\begin{align}
\left\{ \left(
\begin{array}{cccc}
  0 & 0 & 1 & 0 \\
  -1 & 0 & 0 & 0 \\
  0 & 0 & 0 & -1 \\
  0 & 1 & 0 & 0 \\
\end{array}
\right), 
\left(
\begin{array}{cccc}
 \frac12 &
 \frac12&
 \frac12&
 \frac12
\end{array}\right) 
\right\}, 
\left(
\begin{array}{cccccccccccc}
  0 & 0 & 0 & 0 & 0 & 0 & 0 & 0 & 0 & 0 & 0 & 1 \\
  0 & 0 & 0 & 0 & 0 & 0 & 0 & 0 & 0 & 0 & 1 & 0 \\
  0 & 0 & 0 & 0 & 0 & 0 & 0 & 0 & 0 & 1 & 0 & 0 \\
  0 & 0 & 0 & 0 & 0 & 0 & 0 & 0 & 1 & 0 & 0 & 0 \\
  0 & 0 & 0 & 0 & 0 & 0 & 0 & 1 & 0 & 0 & 0 & 0 \\
  0 & 0 & 0 & 0 & 0 & 0 & 1 & 0 & 0 & 0 & 0 & 0 \\
  0 & 0 & 0 & 0 & 0 & 1 & 0 & 0 & 0 & 0 & 0 & 0 \\
  0 & 0 & 0 & 0 & 1 & 0 & 0 & 0 & 0 & 0 & 0 & 0 \\
  0 & 0 & 0 & 1 & 0 & 0 & 0 & 0 & 0 & 0 & 0 & 0 \\
  0 & 0 & 1 & 0 & 0 & 0 & 0 & 0 & 0 & 0 & 0 & 0 \\
  0 & 1 & 0 & 0 & 0 & 0 & 0 & 0 & 0 & 0 & 0 & 0 \\
  1 & 0 & 0 & 0 & 0 & 0 & 0 & 0 & 0 & 0 & 0 & 0 \\
\end{array}
\right)
\end{align}

The Phase-type gates, all make use of non-trivial space group transformations with both $M$ and $b$ non-trivial.
Many of them result in equivalent logical actions. Recording only $7$ independent logical actions which generate the group of cystalline symmetries we have, 
\begin{align}
\left\{ \left(
\begin{array}{cccc}
  0 & 0 & 0 & 1 \\
  0 & 0 & -1 & 0 \\
  0 & -1 & 0 & 0 \\
  1 & 0 & 0 & 0 \\
\end{array}
\right), 
\left(
\begin{array}{cccc}
  \frac12 &
 \frac12 &
 \frac12 &
- \frac12
\end{array}\right) 
\right\},
\left(
\begin{array}{cccccccccccc}
  1 & 0 & 0 & 0 & 0 & 0 & 0 & 0 & 0 & 0 & 0 & 1 \\
  0 & 1 & 0 & 0 & 0 & 0 & 0 & 0 & 0 & 0 & 1 & 0 \\
  0 & 0 & 1 & 0 & 0 & 0 & 0 & 0 & 0 & 1 & 0 & 0 \\
  0 & 0 & 0 & 1 & 0 & 0 & 0 & 0 & 1 & 0 & 0 & 0 \\
  0 & 0 & 0 & 0 & 1 & 0 & 0 & 1 & 0 & 0 & 0 & 0 \\
  0 & 0 & 0 & 0 & 0 & 1 & 1 & 0 & 0 & 0 & 0 & 0 \\
  0 & 0 & 0 & 0 & 0 & 0 & 1 & 0 & 0 & 0 & 0 & 0 \\
  0 & 0 & 0 & 0 & 0 & 0 & 0 & 1 & 0 & 0 & 0 & 0 \\
  0 & 0 & 0 & 0 & 0 & 0 & 0 & 0 & 1 & 0 & 0 & 0 \\
  0 & 0 & 0 & 0 & 0 & 0 & 0 & 0 & 0 & 1 & 0 & 0 \\
  0 & 0 & 0 & 0 & 0 & 0 & 0 & 0 & 0 & 0 & 1 & 0 \\
  0 & 0 & 0 & 0 & 0 & 0 & 0 & 0 & 0 & 0 & 0 & 1 \\
\end{array}
\right)
\end{align}

\begin{align}
\left\{\left(
\begin{array}{cccc}
  0 & 1 & 0 & 0 \\
  1 & 0 & 0 & 0 \\
  0 & 0 & -1 & 0 \\
  0 & 0 & 0 & -1 \\
\end{array}
\right), 
\left(
\begin{array}{cccc}
 \frac12 &
 -\frac12&
 \frac12&
 \frac12
\end{array}\right) 
\right\},
\left(
\begin{array}{cccccccccccc}
  1 & 0 & 0 & 0 & 0 & 0 & 0 & 0 & 0 & 0 & 0 & 1 \\
  0 & 1 & 0 & 0 & 0 & 0 & 0 & 0 & 0 & 0 & 1 & 0 \\
  0 & 0 & 1 & 0 & 0 & 0 & 0 & 0 & 0 & 1 & 0 & 0 \\
  0 & 0 & 0 & 1 & 0 & 0 & 0 & 0 & 1 & 0 & 0 & 1 \\
  0 & 0 & 0 & 0 & 1 & 0 & 0 & 1 & 0 & 0 & 0 & 1 \\
  0 & 0 & 0 & 0 & 0 & 1 & 1 & 0 & 0 & 1 & 1 & 0 \\
  0 & 0 & 0 & 0 & 0 & 0 & 1 & 0 & 0 & 0 & 0 & 0 \\
  0 & 0 & 0 & 0 & 0 & 0 & 0 & 1 & 0 & 0 & 0 & 0 \\
  0 & 0 & 0 & 0 & 0 & 0 & 0 & 0 & 1 & 0 & 0 & 0 \\
  0 & 0 & 0 & 0 & 0 & 0 & 0 & 0 & 0 & 1 & 0 & 0 \\
  0 & 0 & 0 & 0 & 0 & 0 & 0 & 0 & 0 & 0 & 1 & 0 \\
  0 & 0 & 0 & 0 & 0 & 0 & 0 & 0 & 0 & 0 & 0 & 1 \\
\end{array}
\right)
\end{align}

\begin{align}
\left\{ \left(
\begin{array}{cccc}
  0 & 0 & 0 & -1 \\
  0 & -1 & 0 & 0 \\
  0 & 0 & -1 & 0 \\
  -1 & 0 & 0 & 0 \\
\end{array}
\right), 
\left(
\begin{array}{cccc}
 \frac12 &
 \frac12&
 \frac12&
 \frac12
\end{array}\right) 
\right\},
\left(
\begin{array}{cccccccccccc}
  1 & 0 & 0 & 0 & 0 & 0 & 0 & 0 & 0 & 0 & 0 & 1 \\
  0 & 1 & 0 & 0 & 0 & 0 & 0 & 0 & 0 & 0 & 1 & 0 \\
  0 & 0 & 1 & 0 & 0 & 0 & 0 & 0 & 0 & 1 & 0 & 1 \\
  0 & 0 & 0 & 1 & 0 & 0 & 0 & 0 & 1 & 0 & 0 & 0 \\
  0 & 0 & 0 & 0 & 1 & 0 & 0 & 1 & 0 & 0 & 0 & 1 \\
  0 & 0 & 0 & 0 & 0 & 1 & 1 & 0 & 1 & 0 & 1 & 0 \\
  0 & 0 & 0 & 0 & 0 & 0 & 1 & 0 & 0 & 0 & 0 & 0 \\
  0 & 0 & 0 & 0 & 0 & 0 & 0 & 1 & 0 & 0 & 0 & 0 \\
  0 & 0 & 0 & 0 & 0 & 0 & 0 & 0 & 1 & 0 & 0 & 0 \\
  0 & 0 & 0 & 0 & 0 & 0 & 0 & 0 & 0 & 1 & 0 & 0 \\
  0 & 0 & 0 & 0 & 0 & 0 & 0 & 0 & 0 & 0 & 1 & 0 \\
  0 & 0 & 0 & 0 & 0 & 0 & 0 & 0 & 0 & 0 & 0 & 1 \\
\end{array}
\right)
\end{align}

\begin{align}
\left\{\left(
\begin{array}{cccc}
  0 & 0 & 1 & 0 \\
  0 & -1 & 0 & 0 \\
  1 & 0 & 0 & 0 \\
  0 & 0 & 0 & -1 \\
\end{array}
\right), 
\left(
\begin{array}{cccc}
- \frac12 &
 \frac12&
 \frac12&
 \frac12
\end{array}\right) 
\right\},
\left(
\begin{array}{cccccccccccc}
  1 & 0 & 0 & 0 & 0 & 0 & 0 & 0 & 0 & 0 & 0 & 1 \\
  0 & 1 & 0 & 0 & 0 & 0 & 0 & 0 & 0 & 0 & 1 & 0 \\
  0 & 0 & 1 & 0 & 0 & 0 & 0 & 0 & 0 & 1 & 0 & 1 \\
  0 & 0 & 0 & 1 & 0 & 0 & 0 & 0 & 1 & 0 & 0 & 1 \\
  0 & 0 & 0 & 0 & 1 & 0 & 0 & 1 & 0 & 0 & 0 & 0 \\
  0 & 0 & 0 & 0 & 0 & 1 & 1 & 0 & 1 & 1 & 0 & 1 \\
  0 & 0 & 0 & 0 & 0 & 0 & 1 & 0 & 0 & 0 & 0 & 0 \\
  0 & 0 & 0 & 0 & 0 & 0 & 0 & 1 & 0 & 0 & 0 & 0 \\
  0 & 0 & 0 & 0 & 0 & 0 & 0 & 0 & 1 & 0 & 0 & 0 \\
  0 & 0 & 0 & 0 & 0 & 0 & 0 & 0 & 0 & 1 & 0 & 0 \\
  0 & 0 & 0 & 0 & 0 & 0 & 0 & 0 & 0 & 0 & 1 & 0 \\
  0 & 0 & 0 & 0 & 0 & 0 & 0 & 0 & 0 & 0 & 0 & 1 \\
\end{array}
\right)
\end{align}

\begin{align}
\left\{\left(
\begin{array}{cccc}
  0 & 0 & 0 & -1 \\
  0 & 0 & -1 & 0 \\
  0 & -1 & 0 & 0 \\
  -1 & 0 & 0 & 0 \\
\end{array}
\right), 
\left(
\begin{array}{cccc}
 \frac12 &
 \frac12&
 \frac12&
 \frac12
\end{array}\right) 
\right\},
\left(
\begin{array}{cccccccccccc}
  1 & 0 & 0 & 0 & 0 & 0 & 0 & 0 & 0 & 0 & 0 & 1 \\
  0 & 1 & 0 & 0 & 0 & 0 & 0 & 1 & 1 & 1 & 1 & 0 \\
  0 & 0 & 1 & 0 & 0 & 0 & 0 & 1 & 0 & 1 & 0 & 0 \\
  0 & 0 & 0 & 1 & 0 & 0 & 0 & 1 & 1 & 0 & 0 & 0 \\
  0 & 0 & 0 & 0 & 1 & 0 & 0 & 1 & 0 & 0 & 0 & 0 \\
  0 & 0 & 0 & 0 & 0 & 1 & 1 & 0 & 0 & 0 & 0 & 0 \\
  0 & 0 & 0 & 0 & 0 & 0 & 1 & 0 & 0 & 0 & 0 & 0 \\
  0 & 0 & 0 & 0 & 0 & 0 & 0 & 1 & 0 & 0 & 0 & 0 \\
  0 & 0 & 0 & 0 & 0 & 0 & 0 & 0 & 1 & 0 & 0 & 0 \\
  0 & 0 & 0 & 0 & 0 & 0 & 0 & 0 & 0 & 1 & 0 & 0 \\
  0 & 0 & 0 & 0 & 0 & 0 & 0 & 0 & 0 & 0 & 1 & 0 \\
  0 & 0 & 0 & 0 & 0 & 0 & 0 & 0 & 0 & 0 & 0 & 1 \\
\end{array}
\right)
\end{align}

\begin{align}
\left\{\left(
\begin{array}{cccc}
  1 & 0 & 0 & 0 \\
  0 & 1 & 0 & 0 \\
  0 & 0 & 1 & 0 \\
  0 & 0 & 0 & 1 \\
\end{array}
\right), 
\left(
\begin{array}{cccc}
 \frac12 &
- \frac12&
 \frac12&
- \frac12
\end{array}\right) 
\right\},
\left(
\begin{array}{cccccccccccc}
  1 & 0 & 0 & 0 & 0 & 0 & 0 & 0 & 0 & 0 & 0 & 1 \\
  0 & 1 & 0 & 0 & 0 & 0 & 0 & 1 & 1 & 1 & 1 & 0 \\
  0 & 0 & 1 & 0 & 0 & 0 & 0 & 1 & 0 & 1 & 0 & 1 \\
  0 & 0 & 0 & 1 & 0 & 0 & 0 & 1 & 1 & 0 & 0 & 1 \\
  0 & 0 & 0 & 0 & 1 & 0 & 0 & 1 & 0 & 0 & 0 & 0 \\
  0 & 0 & 0 & 0 & 0 & 1 & 1 & 0 & 1 & 1 & 0 & 1 \\
  0 & 0 & 0 & 0 & 0 & 0 & 1 & 0 & 0 & 0 & 0 & 0 \\
  0 & 0 & 0 & 0 & 0 & 0 & 0 & 1 & 0 & 0 & 0 & 0 \\
  0 & 0 & 0 & 0 & 0 & 0 & 0 & 0 & 1 & 0 & 0 & 0 \\
  0 & 0 & 0 & 0 & 0 & 0 & 0 & 0 & 0 & 1 & 0 & 0 \\
  0 & 0 & 0 & 0 & 0 & 0 & 0 & 0 & 0 & 0 & 1 & 0 \\
  0 & 0 & 0 & 0 & 0 & 0 & 0 & 0 & 0 & 0 & 0 & 1 \\
\end{array}
\right)
\end{align}

\begin{align}
\left\{\left(
\begin{array}{cccc}
  1 & 0 & 0 & 0 \\
  0 & 0 & 1 & 0 \\
  0 & 1 & 0 & 0 \\
  0 & 0 & 0 & 1 \\
\end{array}
\right), 
\left(
\begin{array}{cccc}
 \frac12 &
 -\frac12&
 -\frac12&
 \frac12
\end{array}\right) 
\right\},
\left(
\begin{array}{cccccccccccc}
  1 & 0 & 0 & 0 & 0 & 0 & 0 & 1 & 0 & 0 & 0 & 0 \\
  0 & 1 & 0 & 0 & 0 & 0 & 1 & 0 & 0 & 0 & 0 & 0 \\
  0 & 0 & 1 & 0 & 0 & 0 & 0 & 0 & 0 & 1 & 0 & 0 \\
  0 & 0 & 0 & 1 & 0 & 0 & 0 & 0 & 1 & 0 & 0 & 0 \\
  0 & 0 & 0 & 0 & 1 & 0 & 0 & 0 & 0 & 0 & 0 & 1 \\
  0 & 0 & 0 & 0 & 0 & 1 & 0 & 0 & 0 & 0 & 1 & 0 \\
  0 & 0 & 0 & 0 & 0 & 0 & 1 & 0 & 0 & 0 & 0 & 0 \\
  0 & 0 & 0 & 0 & 0 & 0 & 0 & 1 & 0 & 0 & 0 & 0 \\
  0 & 0 & 0 & 0 & 0 & 0 & 0 & 0 & 1 & 0 & 0 & 0 \\
  0 & 0 & 0 & 0 & 0 & 0 & 0 & 0 & 0 & 1 & 0 & 0 \\
  0 & 0 & 0 & 0 & 0 & 0 & 0 & 0 & 0 & 0 & 1 & 0 \\
  0 & 0 & 0 & 0 & 0 & 0 & 0 & 0 & 0 & 0 & 0 & 1 \\
\end{array}
\right)
\end{align}

Together, the Hadamard- and Phase-type gates generate all the permutations found. 
This Clifford group has order $1132462080$, and is a subgroup of the full Clifford group on 6 logical qubits. 
We also note that through a basis change, the above Hadamard-type gate generator will act as $H^{\otimes 4}$ on the first 4 of 6 logicals, and one of the Phase-type gate generators will act as $S^{\otimes 4}$ on the first 4 of 6 logicals.
Therefore giving the full single qubit Clifford group repeated across the first 4 logicals. 
We also have $\text{CZ}_{12}\otimes \text{CZ}_{34}$ and $\text{CZ}_{13}\otimes \text{CZ}_{24}$ on those first four logicals. 

It is also worth noting that since the code is CSS it admits a transversal CNOT between code blocks. 

\section{Surgery}
\label{surgerysection}

To generate additional logical operations, we consider surgery.
This is a generalization of the type of surgery previously considered for two-dimensional toric code patches.
Due to the larger number of logical qubits, 
the effect on the logical qubits is more complicated, 
and due to the potentially complicated HNF, 
it is necessary to construct an appropriate geometry for surgery.
We will first consider the surgery at a topological level,
clarify where the surgery is happening,
and then consider the code distance during the surgery; see also Ref.~\cite{hillmann2024}.

\subsection{Surgery operation on one or two blocks of (2,2) toric code}

First we consider a surgery operation which combines two code blocks into a single code block.
Each code block has $6$ logical qubits and the combined code block also has $6$ logical qubits, 
so the effect of the surgery is to measure $6$ logical operators.

We take two code blocks.  
Topologically, each is some cellulation of a four-torus.  
We then cut each open along some hyperplane of the four-torus.  
Labelling the four directions of a standard four-torus by $1,2,3,4$, 
we pick some direction $i\in \{1,2,3,4\}$, 
and cut open each four-torus at some arbitrarily chosen value of $x_i$, 
exposing two hyperplanes. 
A hyperplane is three-dimensional in this case.  
We then perform surgery, gluing each hyperplane of a given torus 
to a hyperplane of the other torus, 
so that the result is a single four-torus.  
We then reverse this operation 
to split the merger into the original code blocks.

We will now show that indeed exactly $6$ logical operators are measured and we see which ones are measured.  We can identify logical operators by giving both a pair of directions, e.g., $(j,k)$ where $j\neq k$ and also whether it is an $X$-type or $Z$-type logical operator.  
We write such a logical operators in the first code block as $X_{j,k}$ or $Z_{j,k}$, depending on whether it is $X$-type or $Z$-type, and we write the operators in the second code block as $X'_{j,k}$ or $Z'_{j,k}$.
Then, the logical operator can be supported on a plane stretching in the $j$ and $k$ directions, at fixed values of the other two coordinates.  With this notation, an $X$-type logical operator with directions $(1,2)$ will anticommute with a $Z$-type logical operator with directions $(3,4)$, for example.
If $i \neq j$ and $i\neq k$, then the logical operator can be supported on a plane not intersecting the cut, but we may choose, by multiplying by stabilizers, that plane to lie in the hyperplane exposed by the cut.  Then, measurement of the added stabilizers of the code in the larger four-torus will measure the product of the corresponding logical operators in the two code blocks.  That is, for $i=1$ for example, we measure the products $X_{2,3} X'_{2,3}$, $X_{2,4} X'_{2,4}$, $X_{3,4} X'_{3,4}$, 
$Z_{2,3} Z'_{2,3}$, $Z_{2,4} Z'_{2,4}$, $Z_{3,4} Z'_{3,4}$.

On the other hand, no product of operators $X_{1,k}$ or $X'_{1,k}$ or $Z_{1,k}$ or $Z'_{1,k}$ is measured by the measurement of the stabilizers of the larger four-dimensional toric code near the surgery location, because any representative of those operators or of a nontrivial product of them necessarily has some support away from the surgery location.  Using two primes on a logical operator to denote the logical operator on the larger code, we have that $X''_{1,k} = X_{1,k} X'_{1,k}$ and similarly $Z''_{1,k} = Z_{1,k} Z'_{1,k}$, as there are representatives of $X''_{1,k}$ and $Z''_{1,k}$ which may be decomposed in this form.

This thus defines the surgery operation.  
By doing this, we measure exactly the $6$ products of logical operators given above.

If we initialize the second code block to the state 
where all $X$-type logical operators are equal to $+1$, 
 do this surgery 
and finally measure the second code block in the $X$ basis, 
so that we are concerned only with the effect on the $6$ logical qubits in the first code block, 
the effect of the surgery
is to measure $X_{1,k}$ for all $k$.
A similar process may be done with $X$ and $Z$ interchanged.

The same result may be achieved by cutting open a single four-dimensional torus, adding $X$-type or $Z$-type boundary conditions on the given face, and then undoing the operation.  Indeed, in this case it suffices to choose some region near the cut (so some three-dimensional volume) which supports either $X$-type or $Z$-type logical operators $X_{1,k}$ or $Z_{1,k}$, and then measure all qubits in that region in either the $X$ or $Z$ basis.

\subsection{Specifying the hyperplane of surgery}

Suppose each of the two code blocks
is based on a common four-dimensional lattice with Hermite normal form
\[
\begin{pmatrix}
 A_{11} & A_{12} & A_{13} & A_{14} \\
 0 & A_{22} & A_{23} & A_{24} \\
0 & 0 & A_{33} & A_{34} \\
0& 0 & 0 & A_{44}
\end{pmatrix}.
\]
The set of all vertices of a code is~$\ZZ^4 / \Lambda$
where $\Lambda$ is the integer row span of the HNF.
Since we have two codes, we will use $\ZZ^4 / \Lambda$ 
and $\ZZ^4 / \Lambda'$ to distinguish two tori,
even though $\Lambda = \Lambda'$ as subsets of~$\ZZ^4$.
The larger code that is a merger of the two codes
has geometry chosen by picking one of the four rows of the HNF
and doubling all entires in that row.
For example, if we pick the second row, then the resulting Hermite normal form for the larger code 
is then
\[
\begin{pmatrix}
 A_{11} & A_{12} & A_{13} & A_{14} \\
 0 & 2 A_{22} & 2 A_{23} & 2 A_{24} \\
 0 & 0 & A_{33} & A_{34} \\
 0 & 0 & 0 & A_{44}
\end{pmatrix}.
\]
The corresponding integer row span of~$A''$ is $\Lambda''$,
and the larger $4$-torus is $\ZZ^4/ \Lambda''$.
We consider a correspondence of vertices as
\begin{align*}
    \phi :
    \begin{cases}
    \ZZ^4 / \Lambda \ni [(x_1,x_2,x_3,x_4)] &\mapsto  [(x_1,x_2,x_3,x_4)] \in \ZZ^4 / \Lambda'' \\
    \ZZ^4 / \Lambda' \ni [(x'_1,x'_2,x'_3,x'_4)] &\mapsto  [(x'_1,x'_2,x'_3,x'_4) + v]  \in \ZZ^4 / \Lambda''
    \end{cases}
\end{align*}
where $0 \le x^{(')}_i < A_{ii}$
and $v = (0,A_{22},A_{23},A_{24})$ is the chosen row of HNF.
Each qubit at~$x + \frac 1 2 \hat i + \frac 1 2 \hat j$,
where $x \in \ZZ^4 / \Lambda^{(')}$ and $\hat i, \hat j$  
($1 \le i < j \le 4$) 
denote the unit directional vectors,
is mapped to the qubit at~$\phi(x) + \frac 1 2 \hat i + \frac 1 2 \hat j$ of the larger code.

It is clear that all the stabilizers of the smaller codes supported away from the cut
are stabilizers of the larger code,
but those at the cut are replaced by new ones in the larger code during the surgery.
Note that since we are considering tori, rather than a box with open boundary conditions,
there are two locations of the cut:
in the above example where we chose the second row to define a larger code,
one component is at $x''_2 = 0$ and the other is at~$x''_2 = A_{22}$.
These are they hyperplanes of the surgery.

\subsection{Distance under surgery operations}

Consider the two-dimensional case, for comparison, 
doing surgery on a pair of two-tori.
Each logical qubit is identified by only a single direction.
Suppose we are jointly measuring, for example, $Z_1 Z'_1$ and $X_1 X'_1$.
The product $Z_1 Z'_1$ is inferred from the product of $Z$-type stabilizer measurements 
along one location of the cut in the larger torus.
However, there are two such locations as noted above.
So, to create an undetectable logical error, 
one needs two errors in stabilizer measurement.
We call this number, $2$, the ``boundary distance,'' 
denoting the minimal number of errors 
that can lead to an undetectable logical error in the surgery process
if we measured all stabilizers of the larger code once.
So, if each smaller code has distance~$d$, 
it requires $\sim d$ measurement rounds in the larger torus 
to make the boundary distance also equal to~$d$.

Now consider a three-dimensional toric code.
This code is not self-dual.
Let the qubits be on edges, 
and fix the stabilizers to be $Z^{\otimes 4}$ around plaquettes 
and $X^{\otimes 6}$ on vertices.
Then, there are line-like logical $Z$ operators 
and surface-like logical $X$ operators.
So, if the torus is a standard $L$-by-$L$-by-$L$ torus, 
the logical $Z$ operators have minimum weight~$L$, 
while the logical $X$ operators have minimum weight~$L^2$.
We cut the three-torus (two cut locations) 
to create two surfaces which are two-tori and then do surgery.
 
If one measures $Z$-type logical operators, 
either by surgery on two tori or by measuring qubits in the $Z$ basis in a plane,
it takes $\sim L$ errors to make an undetectable logical error
as there are $L$ nonoverlapping representatives in the plane.  
A more general way to compute this number,
which gives a better estimate for more general geometries,
is the following.
Recall that $Z$-type logical measurements in surgery are obtained 
by combining $Z$ stabilizer measurements on plaquettes 
that straddle the two sides of a cut.
Considering only plaquettes in which one direction crosses the cut, 
we can specify a plaquette by one edge in the plane of the cut.
Hence, we can regard errors as \emph{edges} in the cut,
forming a $1$-cochain;
it is a cochain (rather than a chain) 
because a measurement error will be noticeable 
as we deform a logical representative by a plaquette parallel to the cut plane,
so the detection of error is associated with a 2-cell.
For these errors to be undetectable, the cochain must be closed. 
For it to cause a logical error, it must be homologically nonzero.
Finally, there are two different cut surfaces, so the boundary distance 
is twice the minimum length of a $1$-cocycle of a nonzero cohomology class.

On the other hand, it takes only $2$ errors to make an undetectable error in the $X$-type measurement,
as there is just one representative of the $X$ logical operator within the surgery plane.
This is consistent with $X$-type logical operators having higher distance in the bulk: 
a higher distance for logical $X$ errors implies a more resilient logical $Z$ information, 
and hence a larger boundary distance.
 
In summary, if one measures for $\sim L$ rounds on the boundary,
then the boundary $Z$ measurement distance 
is as large as the $X$ distance of the code (i.e., $\sim L^2$) 
and the boundary $X$ measurement distance 
is as large as the $Z$ distance of the code (i.e., $\sim L$).

Finally, we consider surgery on a four-torus.
Now, $Z$-stabilizers are on $3$-cells of the lattice, 
and $Z$-stabilizers crossing the cut correspond to $2$-cells in the hyperplane of the cut.
The boundary distance for $Z$ measurements
is then twice the minimum length 
of a $2$-cocycle of a nonzero cohomology class of the boundary $3$-torus.
 In the standard untwisted $4$-torus, this is~$2L$, linear in $L$.
On the other hand,
$X$-stabilizers are on $1$-cells of the lattice, and those that cross a cut hyperplane are specified by $1$-cells
in the hyperplane of the cut.
(There s no shift in dimension here. 
Perhaps the best way to see this is to think of Poincar\'e duals.
The dual of a $1$-cell relative to the cut hyperplane is a $2$-cell, 
which is the boundary of the dual of the $1$-cell relative to the full $4$-space.)

Therefore,
the boundary distance for $X$ measurement errors 
is then twice the minimum length of a $1$-cycle of nonzero homology of the cut hyperplane.
Both these distances (for the $Z$ and $X$ measurement errors) are equal,
in the particular cellulation of the four-torus that we chose, 
to twice the minimum, over all nonzero vectors in the lattice which lie in the given hyperplane, 
of the $\ell_1$ norm of that vector.

\section{State Injection}
\label{injectionsection}

There are several quantum computing architectures based on CSS codes 
to implement logical Clifford operations to high accuracy, 
and then using state injection to inject ``magic states'' into these codes 
to perform non-Clifford operations.  
In any state injection scheme, 
it is essential to minimize the error created by encoding the state 
(which is initially stored in one or a small number of qubits) into the code.  
For the surface code, a state-of-the-art scheme was given in Ref.~\cite{li2015magic}.  
This scheme can offer reduced error compared to other schemes based on unitary encoding circuits, 
with its error rate potentially \emph{smaller} than that of a single two-qubit gate.

Here we provide a generalization of this scheme, 
valid for an arbitrary stabilizer code ${\cal C}$ with an arbitrary number of qubits.
Application to the four-dimensional toric codes will be an example.
We first consider the case of a CSS code and then describe the case of a non-CSS code.  
We will explain how these schemes can be turned into unencoding schemes.
Let $k$ denote the number of logical qubits.

\subsection{CSS codes}

\noindent
We choose three sets~$S_Z,S_X,U$ of qubits subject to the following properties.
\begin{itemize}
\item[\bf 0.] $U$ has cardinality $k$.
\item[\bf 1.] There is no nontrivial $X$-logical operator supported on~$S_X$.
\item[\bf 2.] There is no nontrivial $Z$-logical operator supported on~$S_Z$.
\end{itemize}
We will soon show that such a choice exists.
Then, the protocol is the following.
We initialize all qubits in $S_Z$ to $\ket 0$ and those in $S_X$ to $\ket +$.
The qubits in~$U$ may be initialized to arbitrary magic states (or even to an entangled state) 
which will be embedded into the logical qubits of the code.
After the initialization, we then measure stabilizers of the code.
One may have to repeat the measurements, 
and the number of rounds required depends on the code; 
for the surface code it may be proportional to the code distance, 
but for single-shot codes it may be $O(1)$.

To see that the protocol produces a code state encoding the state on~$U$,
we recall the following conditions equivalent to~{\bf 1} and {\bf 2} 
by the cleaning lemma and its converse~\cite{bravyi2009no,HaahPreskill}.
\begin{itemize}
\item[\bf 3.] Every $Z$-logical operator 
can be written as a product of stabilizers and an operator supported on $S_Z \cup U$.
\item[\bf 4.] Every $X$-logical operator 
can be written as a product of stabilizers and an operator supported on $S_X \cup U$.
\end{itemize} 
Given these properties, there is a basis of $Z$-type logical operators supported on $S_Z\cup U$ 
and a basis of $X$-type logical operators supported on $S_X\cup U$.
By performing linear algebra on these operators, 
one can construct a basis where for each qubit~$q \in U$ 
there is one $Z$-type logical operator supported on $S_Z\cup \{q\}$ 
and one $X$-type logical operator supported on $S_X\cup \{q\}$.
Call these operators $\tilde Z_q,\tilde X_q$ for qubits $q \in U$.
Then, the protocol embeds the (possibly entangled) state on~$U$ 
into the code subspace, 
with $Z_q \mapsto \tilde Z_q$ and $X_q \mapsto \tilde X_q$.
Note that this gives a proof of the quantum Singleton bound
for CSS codes:
\[
 d - 1 \le \min(|S_Z|,|S_X|) \le \frac{|S_Z|+|S_X|}{2} = \frac{n - k}{2}
\]
where the first inequality follows from the logical operators $\tilde X_q, \tilde Z_q$.

To show that such sets~$S_Z,S_X,U$ exist, 
we start with~$U$ containing all qubits, 
with which {\bf 1} and {\bf 2} are satified, but {\bf 0} is not.
We then remove qubits one at a time from~$U$, 
moving them to either~$S_Z$ or~$S_X$
so that {\bf 1} and {\bf 2} remain true, 
and repeat this until $U$ has cardinality exactly~$k$.

Suppose that $S_Z,S_X,U$ obey {\bf 1-4}.  
Order the physical qubits so that Gauss elimination on $X$-stabilizer group
gives an $X$-stabilizer basis written in the rows of
\[
\begin{pmatrix}
    S_Z & U & S_X \\
    \hline
    A_{l \times s}  & \star_{l \times u} & \star_{l \times t} \\
    0               & B_{l' \times u} & \star_{l' \times t} \\
    0 & 0 & C_{l'' \times t}
\end{pmatrix}
\]
where $s = |S_Z|, u = |U|, t = |S_X|$.
By construction, the submatrix~$A$ has the full rank~$l$.
Let $O$ be a $Z$-logical operator supported on $S_Z \cup U$,
which is general by~{\bf 3},
and we write $O = O_S O_U$ as a product of the tensor components on~$S_Z$ and~$U$.
Clearly, the tensor factor~$O_U$ must commute with all the $X$-stabilizers supported on $U \cup S_X$,
which correspond to the bottom two blocks led by~$B$ and $C$ above.
If we define $G_X$ to be the group of $X$-operators on~$U$ corresponding to the rows of~$B$,
it follows that $O_U$ must commute with all of~$G_X$.
In coding theory language, 
the group~$G_X$ is obtained by shortening the $X$-stabilizer group to~$U\cup S_X$ 
and further puncturing it to~$U$.
Conversely, given any $Z$-operator~$O'_U$ supported on~$U$ 
such that $O'_U$ commutes with all of~$G_X$,
we can find an appropriate $Z$-operator $O'_S$ supported on~$S_Z$
such that the product~$O'_S O'_U$ is a $Z$-logical operator
because $A$ has full rank.
In summary, every $Z$-logical operator~$O$ gives a $Z$-operator $O_U$ that commutes with $G_X$,
and any $Z$-operator that commutes with $G_X$ gives a $Z$-logical operator.

By symmetry, we have $G_Z$ (the puncture of the shortening of the $Z$-stabilizer group).
The two groups $G_X$ and $G_Z$ must commute elementwise
since they are restrictions of commuting groups to a set of qubits of potential overlap.
As before, there are forward and backward maps 
between the set of all $X$-logical operators and the $X$-commutant of~$G_Z$.
Observe that these maps must preserve the commutation relations:
the forward map from $X,Z$-logical operators to its restriction on~$U$
retains all overlapping tensor factors between them,
and
the backward map adjoins tensor factors on~$S_Z,S_X$ where $X,Z$-logical operators do not overlap.
We conclude that a new code~$\mathcal G_U$ defined by a Pauli stabilizer group~$G_X \oplus G_Z$ 
has exactly $k$ logical qubits.

If moving a qubit~$q \in U$ to~$S_X$ violates~{\bf 1},
then $X_q$ times some $X$-type operator on~$S_X$ is a nontrivial $X$-logical operator,
which implies that $X_q$ is a logical operator of~$\mathcal G_U$.
Hence, if a qubit~$q \in U$ could not be moved to $S_Z \cup S_X$,
then the qubit~$q$ fully supports a pair of anticommuting logical operators of~$\mathcal G_U$.
Therefore, whenever~$|U| > k$,
there exists a qubit that can be moved to either~$S_Z$ or~$S_X$
while keeping~{\bf 1} and {\bf 2} valid.
This completes the proof of the existence of~$S_Z,S_X,U$ satisfying {\bf 0}, {\bf 1}, {\bf 2}.

We remark that the protocol can be generalized 
to embed the code space of~$\mathcal G_U$ defined on $m > k$ physical qubits 
into a larger code~$\mathcal C$.

\subsection{non-CSS Codes}

Now assume that ${\cal C}$ is an arbitrary,  non-CSS stabilizer code. 
We will construct four sets of qubits, $S_X,S_Y,S_Z$, and~$U$, 
such that each qubit is in exactly one of these sets.
Our encoding protocols is to initialize qubits in $S_X$ to $\ket + = X \ket +$, 
those in $S_Y$ to $\ket i = Y\ket i$, 
and those in $S_Z$ to $\ket 0 = Z\ket 0$,
and then measure stabilizers.
Let us write $S=S_X \cup S_Y \cup S_Z$.  
We will say that a Pauli operator~$P$ is ``standard'' 
if every factor of~$P$ on~$S_X$ is~$X$,
every factor on~$S_Y$ is~$Y$,
and every factor on~$S_Z$ is~$Z$.
A standard operator may act on~$U$ arbitrarily.
We will require that
\begin{itemize}
 \item[\bf a.] There is no nontrivial standard logical operator supported on~$S$.
 
 \item[\bf b.] Every logical Pauli operator is a standard operator times a stabilizer.
 
 \item[\bf c.] $|U| = k$.
\end{itemize}
With these requirements,
we find $k$ pairs of stadard logical operators $\tilde X_q$ and $\tilde Z_q$
where $\tilde X_q$ is Pauli~$X_q$ on a qubit~$q \in U$ multiplied by a standard operator supported on~$S$,
and $\tilde Z_q$ is Pauli~$Z_q$ on~$q$ multiplied by a standard operator supported on~$S$.
It is clear that the protocol injects the state on~$U$
into the code space of~$\mathcal C$
since $\tilde X_q, \tilde Z_q$ commute with all the measurements.
The existence proof for the partition $U \cup S_X \cup S_Y \cup S_Z$
will be constructive and similar to that of the CSS case.

We first note that {\bf b} is equivalent to~{\bf a}.
Since every standard operator on~$S$ commutes with any other standard operator on~$S$,
any standard logical operator supported on~$S$ must be trivial,
proving {\bf b} $\Rightarrow$ {\bf a}.
For {\bf a} $\Rightarrow$ {\bf b},
we consider the evolution of the stabilizer group
when measuring all individual qubits in~$S_W$ in the $W$ basis, where $W = X,Y,Z$.
By~{\bf a}, these measurements do not reveal any logical information,
and therefore each measurement outcome here is independent of the underlying logical state.
This is only possible if every logical operator of~$\mathcal C$ 
can be multiplied by a stabilizer so that it commutes with the measurement operator,
which means~{\bf b}.

Suppose that we have found some $S = S_X \cup S_Y \cup S_Z$ and~$U$ satisfying~{\bf b},
but not~{\bf c}.
Consider a linear equation for standard logical operators~$O$,
derived by commutation relations to stabilizers:
\[
    \begin{pmatrix}
        S & U \\
        \hline
        A & \star \\
        0 & B
    \end{pmatrix} 
    \begin{pmatrix} [O_S] \\ [O_U] \end{pmatrix} = 0
\]
where $[O_{S}]$ and $[O_U]$ are bit vectors 
corresponding to the tensor factors~$O_S$ on~$S$ and $O_U$ on~$U$ 
of a standard Pauli operator~$O = O_S O_U$.
The full operator~$O$ is typically represented by~$2n$-bit vector,
but since we are considering standard operators whose factor on~$S$ is constrained,
the component~$[O_S]$ has $|S|$ bits, while $[O_U]$ has $2|U|$ bits.
We may assume that this linear equation is in a row echelon form,
implying that the submatrix~$A$ has rank equal to the number of rows of~$A$.
The rows below~$A$ (the rows of $B$ padded by zeros in front) 
correspond to the stabilizers that commute with every standard operator supported on~$S$,
which are precisely the standard stabilizers.
Let $G$ be the group generated by the tensor factors on~$U$ of all standard stabilizers;
every element~$g \in G$ has a standard operator~$s$ supported on~$S$ 
such that $gs$ is a stabilizer of~$\mathcal C$.
The group~$G$ is abelian since the commutation relation
between two standard operators is determined by that of the factors on~$U$ only.
The abelian group~$G$ defines a new Pauli stabilizer code~$\mathcal G_U$ on~$U$.

Every standard logical operator of~$\mathcal C$ restricts 
to a logical operator of~$\mathcal G_U$,
and every logical operator of~$\mathcal G_U$
gives rise to a standard logical operator of~$\mathcal C$ because $A$ has full rank.
This correspondence preserves the commutation relation,
and therefore, by~{\bf b}, the code~$\mathcal G_U$ has exactly $k$ logical qubits.
Now suppose that moving a qubit~$q \in U$ to~$S_X$ violates~{\bf a}.
This means that $X_q$ times a standard operator on~$S$ 
is a nontrivial logical operator of~$\mathcal C$,
which in turn means that $X_q$ is a nontrivial logical operator of~$\mathcal G_U$.
Hence, if every qubit of~$U$ could not be moved to~$S$,
then every qubit of~$U$ would fully support a triple of nontrivial logical operators,
and the number of logical qubits of~$\mathcal G_U$ would be $|U| > k$.
Therefore, some qubit of~$U$ can be moved to~$S$ while keeping {\bf a} true, whenever~$|U| > k$.

\subsection{Unencoding scheme}

The schemes above inject (or encode) $k$ qubits into a code.
Conversely, these schemes can be used for the inverse operation of the encoding, 
mapping the logical qubits of a code into physical qubits.
Simply take an arbitrary stabilizer code, 
and measure qubits in~$S_X$ in the $X$ basis, 
those in~$S_Y$ in the $Y$ basis, and those in~$S_Z$ in the $Z$ basis.  
Then, $U$~holds the state of the logical qubits, 
up to a Pauli frame change 
which can be computed from the measurement outcomes.
This was alluded in the argument for the equivalence between~{\bf a} and {\bf b} above.





\section{Slicing and Twisting}
\label{slicetwist}
Now we give a more general discussion of slicing and twisting, beyond the context of toric codes.  This has two uses.  (1) After slicing, we can produce more complicated stabilizer states of logical qubits, rather than just Bell pairs or GHZ states.  (2) After slicing, we can encode the logical qubits into arbitrary codes.  In the case of more general codes, the rotated boundary conditions of the three-dimensional toric code (used to improve distance) get replaced by more general automorphisms of the code.

Let $D$ be some CSS quantum code.  
In our examples above with a 3-torus, 
$D$ will be a 2-dimensional toric code.  
We identify this code $D$ with a chain complex with $\mathbb{Z}_2$ coefficients.
This chain complex has three degrees, labeled degree $0,1,2$, 
and we identify $X$-stabilizers, qubits, and $Z$-stabilizers with basis elements for vector spaces of degree $0,1,2$ respectively.  
The commutation of the $Z$- and $X$-stabilizers is guaranteed by the fact that the square of the boundary operator in a chain complex is $0$.  
Let $\partial_D$ be the boundary operator for this chain complex $D$.  Let $D_0,D_1,D_2$ be the vector spaces of degree $0,1,2$ in this chain complex, respectively.

Let $C$ be some chain complex with $\mathbb{Z}_2$ coefficients with only two degrees, labelled degree $0,1$.  We may identify this chain complex with a classical code.  In our examples above with a three torus, $C$ is a \emph{one-dimensional} toric code on a circle.  We refer to basis elements of degree $j$ as $j$-cells, and $C$ has $\nslice$ $0$-cells and $\nslice$ $1$-cells.  Let $\partial_C$ be the boundary operator for $C$.
Let $C_0,C_1$ be the vector spaces of degree $0,1$ in this chain complex, respectively.

Assume that $D$ admits some automorphism group.  
These automorphisms of $D$ will be permutations of the $0$-cells, $1$-cells, and $2$-cells of $D$ (mapping $j$-cells to $j$-cells for $j=0,1,2$) which commute with the boundary operator of~$D$.
We use symbols such as $\phi$ to denote automorphisms of $D$.
In the case of the examples above, the automorphisms that we use are translations of the two-dimensional toric code.
Some two-dimensional toric codes may also admit other automorphisms such as reflections or rotations, 
but we do not use those in the above examples.

We now take a twisted product of $C$ and $D$.  This is the same twisted product used in \cite{hastings2021fiber}, and the twisted product is taken for the same reason, to improve code distance.  We first recall the untwisted product, or ``homological product"\cite{bravyi2014homological}.
The chain complex $C\otimes D$ has degrees $0,1,2,3$.  The vector space of degree $k$ in $C\otimes D$ is
\[
\bigoplus_{0\leq j \leq k} C_j \otimes D_{k-j}\, .
\]
There is an obvious preferred basis that we use for these vector spaces; 
the basis elements for the vector space of degree $k$ are given by 
a pair of a basis element of degree $j$ in $C$ and a basis element of degree $k-j$ in $D$.
The boundary operator of the untwisted product is
\[
\partial_C \otimes I + I \otimes \partial_D.
\]
If we had used coefficients other than $\mathbb{Z}_2$, it would be necessary to include signs in this boundary operator.

The twisted product has the same vector space of degree $k$ and same basis as $C\otimes D$ does.  However, the boundary operator is different.  Let $c^0,c^1$ denote $0$- and $1$-cells in $C$, respectively.  Let $d$ denote an arbitrary chain in $D$.  Then define the twisted boundary operator by
\[
\partial (c^0\otimes d)=c^0 \otimes \partial d
\]
and
\[
\partial (c^1 \otimes d)=c^1 \otimes \partial d + \sum_{a^0\in \partial c^1} a^0 \otimes \phi(c^1,a^0) d.
\]
That is, the boundary operator acting on $c^0\otimes d$
is the same as in the untwisted case, 
but acting on $c^1 \otimes d$ the boundary operator is different.
Here $\phi(c^1,a^0)$ is an arbitrary choice of automorphism of $D$ for each pair $c^1,a^0$, where $a^0$ is a $0$-cell of $C$.

In the (twisted) product, we identify $0$-, $1$-, $2$-cells with $X$-stabilizers, qubits, and $Z$-stabilizers, respectively.  $3$-cells correspond to redundancies of the $Z$-stabilizers, enabling a single shot property.

One may verify that the examples above do indeed correspond to this twisted product.  
The elements right of the diagonal of the first row in the Hermite normal form 
correspond to the automorphism $\phi$ of the two-dimensional toric code used.
Now we consider the case of general $C,D$.  We again prepare all physical qubits in the $X=+1$ state.  We then measure $Z$-stabilizers of the twisted product.  We then measure out all qubits corresponding to $1$-cells of the form $c^1 \otimes d^0$; this is the generalization of measuring qubits corresponding to edges in a given direction.

The remaining qubits are on $1$-cells of the form $c^0 \otimes d^1$.  Thus, we have one quantum code $D$ for each $0$-cell of $C$.  One may verify that, for each $0$-cell $c^0$ of $C$,  the resulting state on cells $c^0 \otimes d^1$ is indeed an eigenstate of each stabilizer of $D$, possibly with a sign depending on measurement outcome.

Let us now analyze the state of the logical qubits of these codes~$D$.
First we consider the untwisted case.
Logical $X$-type operators correspond to elements of first cohomology,
and the  K\"{u}nneth formula gives us representatives for these elements.
These represenatatives are of the form
\[
    \tilde c^j \otimes \tilde d^{1-j},
\]
where $\tilde c^j$ is a representative of $j$-th cohomology for $C$ and $\tilde d^{1-j}$ is a representative of $(1-j)$-th cohomology for~$D$.%
\footnote{Here we are denoting a cochain (a linear functional on chains) by a chain (e.g., $\tilde c^j$)
under an obvious rule that a chain~$c$ as a cochain evaluated on a chain~$c'$ assumes~$1 \in \ZZ_2$
if and only if $c$ and $c'$ as sets of cells overlaps over an odd number of cells.}
The operators corresponding to elements $j=1$ are measured out, so the remaining operators correspond to elements of the form $\tilde c^0 \otimes \tilde d^1$.  Each representative $\tilde d^1$ corresponds to an $X$-type logical operator of $D$.  Denote the $X$-type logical operators of $D$ by $\tilde X^a$, where $a$ is some discrete index.  We have several copies of $D$; denote the $X$-type logical operator $\tilde X^a$ on a given copy $v$ by $\tilde X^a_v$.
Then we have logical stabilizers of the form $\prod_{v \in \tilde c^0} \tilde X^a_v$, for each representative $c^0$.  For example, in the case of a GHZ state from slicing a three-dimensional toric code, these are the operators $\prod_i \tilde X_i^a$.

Since the entangled logical state is pure, we can deduce that we also have $Z$-type logical stabilizers corresponding to the commutant of these $X$-type logical operators.  
One may also calculate this directly.
Let $\tilde d^{1}$ be a representative of first homology for $D$.
For any $1$-cell $c^1$ we have
$\partial(c^1 \otimes \tilde d^1)=(\partial c^1) \otimes \tilde d^1$,
and these correspond to logical $Z$-type stabilizers.
If $\tilde d^1$ corresponds to some logical operator $\tilde Z^a$,
then we have $Z$-type logical stabilizers
$\prod_{v \in \partial c_1} \tilde Z^a_v$.
Indeed, this is the commutant of the $X$-type logical operators.
Thus, we the logical qubits on copies of $D$ can be put into  any pure transverse CSS stabilizer state such that all stabilizers have eigenvalue $+1$\footnote{We define a transverse CSS stabilizer state for a code with $k$ logical qubits to be a $k$-fold tensor product of a CSS stabilizer state with the tensor product taken in the obvious way, i.e., if all stabilizers have eigenvalue $+1$ then this is any state that can be created by initializing each code block with all logical qubits in the same state, either $0$ or $+$, and then applying transversal CNOT gates.}.

Thus far, we have considered the untwisted case.
In the twisted case, if the twist \emph{acts trivially on homology and cohomology}, 
so that it maps a representative of a given class to a representative of the same class,
then the entanglement of the logical operators is the same as in the untwisted case.
The K\"{u}nneth formula representatives
$\tilde c^0 \otimes \tilde d^{1}$ need not be elements of the cohomology of the twisted product; 
indeed, they need not be coclosed in this case.
However, we can make them coclosed by adding appropriate cells of the form $c^1 \otimes d^0$ 
to cancel terms arising from the twist.
Similarly, for the logical $Z$-type stabilizers, 
we no longer have
$\partial(c^1 \otimes \tilde d^1)=(\partial c^1) \otimes \tilde d^1$, 
but rather some automorphism is applied to $\tilde d^1$.
However, if the automorphism acts trivially on homology, 
then we have the same stabilizers for the logical operators.
If the twist does not act trivially on homology and cohomology, 
it is more complicated and we do not consider this case.

\section{Asymptotics in Large Number of Dimensions}
\label{higherdim}

\Cref{geometrymethod} has given specific examples of rotated lattices for dimensions $D=3,4$.  In this case, while rotation leads to some improvement in the ratio of $n/d^2$ in the resulting quantum code, the asymptotic performance, at large code distance $d$, is the same, up to constant factors, as the unrotated code.
We now consider arbitrary dimension $D$, and consider the asymptotic behavior in dimension $D$.  First we give an explicit family for $D$ a power of $2$, and then we give a randomized construction which is optimal up to constant factors.
Our discussion will focus solely on the minimum $\ell_1$ distance in the lattice, rather than considering the code distance for a $(D/2,D/2)$ toric code, which is much more complicated\cite{hastings2017quantum}.

\subsection{Explicit family}
Let
\begin{align}
    H &\equiv \begin{pmatrix}
        1 & 1 \\
        1 & -1
    \end{pmatrix},\\
    H^2 &= 2 I_2.\\
    \det H &= 2.
\end{align}
Consider $D = 2^t$ dimensions, and take $H^{\otimes t}$.  
This matrix is a so-called Hadamard matrix, sometimes denoted by~$H_D$.
The determinant is~$2^{t 2^{t-1}}$  because $\det[ (H^{\otimes t})^2 ] = \det (2^t I_{2^t}) = 2^{t 2^t}$.

In $D=4$ ($t=2$), the minimum nonzero $\ell_1$-norm of~$\Lambda$ is~$4$.
Hence, the rotated $(1,3)$-toric code has parameters $[[4 \cdot 16 \ell^4, 4, 4\ell]]$ with the smallest in the family being $[[64,4,4]]$.

Note that we have
\begin{align}
    H_D &=\begin{pmatrix}
        H_{D/2} & H_{D/2} \\ H_{D/2} & -H_{D/2}
    \end{pmatrix} 
    \sim
    \begin{pmatrix}
        H_{D/2} & H_{D/2} \\ 0 & -2H_{D/2}
    \end{pmatrix} \, .
\end{align}
From this, we prove inductively that the $\ell_1$ systole is equal to $D$.
Note first that each row has $\ell_1$ norm $D$ so that the $\ell_1$-systole is at most $D$.
Now, assume that $H_D/2$ has $\ell_1$ systole equal to $D/2$.
Consider a nonzero vector $(u, u - 2v)$ in the lattice defined by the rows of~$H_D$,
where $u,v$ are in the row span of~$H_{D/2}$.
If $u = 0$, then $v \neq 0$ and the $\ell_1$ norm is $\ge 2 \cdot D/2 = D$.
If $u \neq 0$ but $u - 2v = 0$, then $u = 2 v \neq 0$ and the $\ell_1$-norm is $\ge 2 \cdot D/2 = D$.
If $u \neq 0$ and $u - 2v \neq 0$, then $\norm{(u, u-2v)}_1 = \norm{u}_1 + \norm{u-2v}_1 \ge D/2 + D/2 = D$.

\subsection{The ``$\ell_1$ Hermite constant"}
Now consider a $D$-dimensional lattice $L$, for arbitrary $D$.  How large can the ratio
$$\frac{\sys(L)}{\det(L)^{1/D}}$$ be?
If we define $\systwo(L)$ to be the shortest length with the $\ell_2$ norm of a nonzero vector in an $D$-dimensional lattice $L$, then it is well known (see references in \cite{nguyen2009hermite} for example) that
$${\rm sup}_{L,\det(L)\neq 0} \frac{\systwo(L)}{\det(L)^{1/D}}=\Theta(\sqrt{D}),$$
where we take the supremum over all $D$-dimensional lattices $L$.  The square of this ratio is called Hermite's constant.

\begin{theorem}
For a $D$-dimensional lattice $L$, we have
    ${\rm sup}_{L,\det(L)\neq 0} \frac{\sys(L)}{\det(L)^{1/D}}=\Theta(D)$.
\end{theorem}
\begin{proof}
The upper bound ${\rm sup}_{L,\det(L)\neq 0} \frac{\sys(L)}{\det(L)^{1/D}}=O(D)$ follows immediately from the $\ell_2$ bound
${\rm sup}_{L,\det(L)\neq 0} \frac{\systwo(L)}{\det(L)^{1/D}}=O(\sqrt{D})$ combined with the fact that
the $\ell_1$ norm of a vector is at most $\sqrt{D}$ larger than the $\ell_2$ norm of that vector.
So, we have to prove the lower bound ${\rm sup}_{L,\det(L)\neq 0} \frac{\sys(L)}{\det(L)^{1/D}}=\Omega(D)$.
Our proof will leave some constants arbitrary; the reader may be interested to optimize the choices of these constants to obtain the best bound using the methods here.  However, it is possible that even better lattices could be found by other techniques.

Pick some constant $C>1$.  
Let $p$ be the least prime larger than $C^D$.
Consider a lattice $L_A$ defined by a Hermite normal form
\[
A=\begin{pmatrix}1 &&&& a_1 \\ &1&&& a_2 \\ &&1&&a_3 \\ &&&\ddots&\vdots \\ &&&& p
\end{pmatrix}.
\]
Choose the entries $a_1,a_2,...,a_{D-1}$ randomly from $0,\ldots,p-1$, uniformly and independently.
Note that $\det(L)^{1/D} = C (1+o(1))$. 

We use a first moment method to show that, with high probability, $\sys(L_A)=\Omega(D)$, 
by bounding the average (over choices of~$a_i$) number of nonzero vectors $v$ in the lattice
with $\ell_1$ norm smaller than $rD$, for some given~$r > 0$.
We will show that, for sufficiently small $r$ this number is~$o(1)$,
and so with high probability, $L_A$ has $\sys(L_A)\geq rD$, proving the theorem.

Such a vector $v$ is given by $v=uA$ for a nonzero integer vector $u$.  
It is immediate that at least one of the first $D-1$ entries of~$u$ must be nonzero.
Further, $$|v|_1 \geq \sum_{i=1}^{D-1} |u_i|.$$

We have $v_D=\sum_{i=1}^{D-1} u_i a_i \bmod p$,
and so for a given choice of $u_1,\ldots,u_{D-1}$ with at least one $u_i$ nonzero, 
and a random choice of $a_1,\ldots,a_{D-1}$, 
we have $v_n$ equal (mod $p$) to a uniformly random number in the range $0,\ldots,p-1$.
The probability that this number is within $rD$ of $0$ (mod $p$) is $(2rD+1)/p$.

So, the average number of nonzero vectors in the lattice
with $\ell_1$ norm smaller than $rn$ is bounded by \[ N_{rD} \frac{2rD+1}{p} = O\left(\frac D {C^D}\right),\]
where $N_{rD}$ is the number of choices of $u_1,\ldots,u_{D-1}$ such that $\sum_{i=1}^{D-1} |u_i| \leq rD$.

Bounding $N_{rD}$ is a standard computation and we have
$$N_{rn}=O(\exp(C'(r) D)),$$
where $C'(r)$ is a constant depending on~$r$ with~$\lim_{r \to 0} C'(r) = 0$.
So, for any given $C>1$, one may choose $r$ sufficiently small that
$C'(r) < C$ and so $N_{rD} \frac{rD}{p} = o(1)$.
\end{proof}

\section{24-Cell Honeycomb and Its Subdivisions}
\label{24cell}
The four-dimensional hypercubic lattice leads to stabilizers of weight $6$, with each qubit participating in $4$ stabilizers.  It is interesting to ask whether these numbers can be reduced by a different cellulation of the torus

Here we give an example that makes some progress: it reduces the weight of one type of stabilizer without increasing the weight of the other type on average.

The starting point is a cellulation of ${\mathbb R}^4$ called the $24$-cell honeycomb.  We will review some of the geometry of this below to guide the reader.  The incidence matrix of this is found at \cite{bendwavy}.

The cellulation has four-dimensional cells which are convex polytopes called $24$-cells. The terminology is perhaps confusing; many authors use $d$-cell to refer to a $d$-dimensional cell but here a $24$-cell is a four-dimensional cell.

The three-dimensional cells are octahedra.  The two dimensional cells are triangles.  Thus, taking $X$ stabilizers on octahedra and $Z$-stabilizers on edges,
each $X$-stabilizer has weight $8$.  At the same time, each $Z$ stabilizer has weight only $4$.  Each qubit on a triangle participates in $3$ $X$-stabilizers and $3$ $Z$-stabilizers.

Thus far, while this cellulation reduces the $Z$-stabilizer weight compared to the hypercubic lattice, it increases the $X$-stabilizer weight.  However, we now subdivide each octahedron into two solid pyramids, with each pyramid having a four-sided basis.  Thus each pyramid corresponds to an $X$ stabilizer of weight $5$, so the weight is reduced below $6$.
Now, there are qubits on triangles and squares, so some qubits are in $3$ $Z$ stabilizers and some are in $4$.

The $X$-stabilizer weight increases, however.  Each octahedron has $12$ edges and we have attached an additional square to $4$ of those edges, increasing the weight of the corresponding $X$ stabilizer by $1$.  Thus, when we subdivide a given octahedron, we increase the weight of $1/3$ of the edges in that octahedron by $1$.  Each edge is in $6$ octahedra, so on average, the weight of an $X$-stabilizer is increased by $6/3=2$, and hence the average weight of $X$ stabilizers is equal to $6$ if we subdivide even octahedron; the precise pattern of weights will depend on how we subdivide them.

Let us return, however, to to considering the $24$-cell honeycomb, without subdivision.
The $24$-cell honeycomb gives a cellulation of $\mathbb{R}^4$.  Instead, we want some finite number of $24$-cells which will cellulate ${\mathbb T}^4$.
What follows is essentially review.

Indeed, let us consider a single $2$-cell.
The $24$-cell has $24$ vertices, which conveniently may be choosen to be any of the ${4 \choose 2}=6$ permutation of $(\pm 1,\pm 1,0,0)$.
The edges have length $\sqrt{2}$, where each edge goes from a vertex with two nonzero coordinates to some other vertex with two nonzero coordinates, where the two vertices share a single nonzero coordinate.
There are two types of triangles.   In one type, there is one coordinate which is nonzero in all three vertices, while the other coordinates are nonzero in exactly one vertex.
There are $4*2^4=64$ such triangles, as we pick one of the $4$ coordinates, and pick $4$ signs.
In the other type, there is one coordinate which is zero in all three vertices.  There are $4*2^3=32$ such triangles, as we pick a coordinate to be zero, and pick $3$ signs.  This gives a total of $96$ triangles.

There are two types of octahedra.  In one type, we pick some one of the four coordinates, and in every vertex that coordinate is $+1$ or $-1$ (the same value in all vertices).  In the other $3$ coordinates, there is one vertex where that coordinate is $+1$ and one where it is $-1$, and it is zero in all the others. There are  $8$ such octahedra as we pick a coordinate and a sign.

Also, there are octahedra with vertices $$(s_1,s_2,0,0),(s_1,0,s_3,0),(s_1,0,0,s_4),(0,s_2,s_3,0),(0,s_2,0,s_4),(0,0,s_3,s_4),$$ where each $s_1,s_2,s_3,s_4$ is a sign $\pm 1$.  There are $2^4=16$ such octahedra.
This gives a total of $24$ octahedra, hence the name $24$-cell.

The $24$-cell honeycomb is the Voronoi cellulation of the $D_4$ lattice.  Let us see how this arises.  The centers of the octahedra are readily calculated.
For the first type of octahedron, the center is at some permutation of $(\pm 1,0,0,0)$, while for the second type the center is at $(s_1/2,s_2/2,s_3/2,s_4/2)$.
The centers of these octahedra are at the midpoint between two vertices of the $D_4$ lattice.  Thus, doubling these centers, we can find a vector going from one vertex of the $D_4$ lattice to neighboring one, i.e., the given by permutation of $(\pm 2,0,0,0)$ and $(s_1,s_2,s_3,s_4)$.

From these vectors going from a vertex of the $D_4$ lattice to a neighboring one, we can find a basis for the $D_4$ lattice, being
$(2,0,0,0)$, $(0,2,0,0)$, $(0,0,2,0)$, $(1,1,1,1)$.  In fact, this choice of lattice basis vectors is conventionally used for the $F_4$ lattice, which is the same as the $D_4$ lattice up to rotation of ${\mathbb R}^4$.  The reader may verify that by rotating and rescaling these can be turned into the basis
$(1,1,0,0),(1,0,1,0)$,(1,0,0,1)$, (2,0,0,0)$ of the $D_4$ lattice.

We now see how to take a cellulation of ${\mathbb T}_4$.  Simply take the cellulation of ${\mathbb R}_4$ and mod out by some sublattice of the $D_4$ lattice.  The smallest code is given by modding out by the lattice itself.  In this case, several of the $96$ triangles become identified with each other and there are only $32$ triangles total.  Indeed, there are $24$ edges, $32$ triangles, and $12$ octahedra per lattice vertex in general.

Thus, we can define a code on $32$ qubits from the minimal cellulation.
A numerical search verifies that the minimum weight $X$-type logical operator has weight $6$.  Unfortunately, the minimum weight $Z$-type logical operator has weight only $2$.  Likely if we subdivide the octrahedra, the $Z$ distance will increase.  However, it is unlikely to be a useful code in this minimal example.  Larger cellulations may lead to useful codes with reduced stabilizer weight compared to $(2,2)$ toric codes on hypercubic lattices.

\section{Numerical simulations}
\label{numsim}
In this section we discuss some numerical tricks we used to study these codes.
Using the Hermite normal form, it is possible to exhaustively enumerate lattices and find the minimum determinant $\det(L)$ for a given distance $\sys(L)$.
To do this quickly on a computer, one can iterate over all possible diagonal elements whose product is less than or equal to some guessed upper bound on the determinant.  If the product of these diagonal elements is less than or equal to the best determinant found so far on the search, one then iterates over off-diagonal elements.  One can quickly discard many of these lattices, as the sum of entries on each row must be at least the target $\sys$.
If the given Hermite normal form passes these tests, one can run a further search to verify the distance $\sys$.

Several tricks can be used to speed up the enumeration used to determine $(2,2)$ distance.  
We label the qubits $0,1,2,...,q_{max}$ for some $q_{max}$.  For these $(2,2)$ codes, $q_{max}=6\det(L)-1$.
The plaquettes can be labelled by vertices of the lattice and by a choice of two lattice directions.

Then, we use a recursive routine with
roughly the following pseudocode.  This is a recursive routine {\bf check}($S$,$w_{max}$,$q_{start}$) defined by:
\begin{itemize}
\item[1.] {\bf for} $q=q_{start}$ {\bf to} $q_{max}$ {\bf do}:
\begin{itemize}
\item[A.] Let $S_{new}$ equal $S$ exclusive-OR with the pattern of stabilizers violated by a single $Z$ error on qubit $q$.

\item[B.] {\bf if} $S_{new}=0$, then an error pattern that produces no violated stabilizers has been found.  See below.

\item[C.] {\bf if} $|S_{new}|\leq 4*(w_{max}-1)$, call {\bf check}($S_{new}$,$w_{max}-1$,$q+1$).
\end{itemize}
\end{itemize}

This is a recursive routine to enumerate all patterns of $Z$ errors on qubits with weight\footnote{The weight of some pattern of errors is simply the number of errors in the pattern.}  at least $1$ and at most $w_{max}$ for some $w_{max}$, with all qubits taken in the range $q_{start},\ldots,q_{max}$ which produces some given syndrome $S$.  By calling this routine with the empty syndrome $S=0$ (i.e., no violated stabilizers) at with $q_{start}=0$ we can find all  error patterns which violate no stabilizers with weight at most some $w_{max}$.  Then, one may test which of these produce a logical error by computing their $\mathbb{F}_2$ inner product with some known representatives of $X$-type logical operator.

A basis of these logical operator representatives is particularly easy to write down if $\det(L)$ is odd.  For each of the ${4 \choose 2}=6$ different choices of two lattice directions, one takes the sum over all plaquettes with that given choice.

This performs the desired enumeration.  Step {\bf 1C} contains two tricks.  First, note that ${\bf check}$ is called with $q_{start}=q+1$, so that when calling ${\bf check}$ with some given $w_{max}$ we only enumerate over ${q_{max} \choose w_{max}}$ different errors patterns.  Second, we only call recursively check if 
$|S_{new}|\leq 4(w_{max}-1)$, where $|S_{new}|$ is the Hamming weight; this is done because a single qubit error only changes the value of $4$ stabilizers, so if this inequality does not hold, it is not possible for the recursive call to ever give a pattern with no violated stabilizers.  Use of this inequality leads to a large speedup.

In step {\bf 1B}, we have simply indicated in the pseudocode that we have found an error pattern.  This error pattern is given by concatenating $q$ with whatever error pattern was chosen in the previous calls of check.  One may keep track of this error pattern found on previous then by adding some additional variables, or one may simply keep track of the inner product of the error pattern with various $X$-logical operator representatives and update that inner product on the recursive call, to determine if there is a nontrivial logical error.

This can be sped up further using bitwise manipulations.
If there are $64$ or fewer stabilizers, one can store which stabilizers are violated as a $64$ bit integer, with each bit corresponding to a given stabilizer.  With more than $64$ such stabilizers, one can store which stabilizers are violated as an array of such integers.  One stores an array of $64$ bit integers (or an array of arrays if there are more than $64$ stabilizers), of size $q_{max}$.  Each entry of the array contains the a bit string (or array of bit strings) showing the stabilizer pattern caused by a single qubit error.  Then, step {\bf 1A} can be done using bitwise-XOR, and the weight in {\bf 1C} can be computed using builtins such as popcount.  Similarly, one can keep track of the inner product of an error pattern with $X$-logical operator representatives using bitwise manipulations.

A further speedup using translation symmetry is useful for larger sizes.  Note that given any logical operator, any translate is also a logical operator.  So, we may assume that some error occurs on one of the $6$ plaquettes corresponding to some fixed vertex.  It is convenient to take some ordering of the qubits so that these plaquettes correspond to qubit labels $0,\det(L),2\det(L),3\det(L),4\det(L),5\det(L)$.  So, we may use the following algorithm to find logical operators of weight at most $w_{max}$.  This is a loop which then calls the recursive routine {\bf check}.

\begin{itemize}
\item[1.] {\bf for} $q$ in  $0,\det(L),2\det(L),3\det(L),4\det(L),5\det(L)$ {\bf do}:
\begin{itemize}
\item[A.] Let $S_{new}$ be the pattern of stabilizers violated by a single $Z$ error on qubit $q$.

\item[B.] Call {\bf check}($S_{new}$,$w_{max}-1$,$q+1$).
\end{itemize}
\end{itemize}

This will enumerate logical operators up to translation symmetry.  To count the actual number of logical operators, one must consider translates of the operators found by this routine.  The number of translates of a given operator is at most $\det(L)$ but may be smaller.

Using these tricks, we were able to obtain the results in \cref{disttable4d} in a reasonable time on a single core.
One may be able to further speedup the enumeration for even larger examples using the fact that one can assume that any minimum weight logical operator has connected support, but this makes the enumeration more complicated.

\bibliography{references}

\noindent$^*$ Microsoft Quantum. \\ $^\dagger$ Current address, Stanford University and Google Quantum AI, all research performed at Microsoft
\end{document}